# Network Topologies for Composable Data Centres

Opeyemi O. Ajibola, Taisir E. H. El-Gorashi, and Jaafar M. H. Elmirghani, *Fellow, IEEE*

***Abstract*—Suitable composable data center networks (DCNs) are essential to support the disaggregation of compute components in highly efficient next generation data centers (DCs). However, designing such composable DCNs can be challenging. A composable DCN that adopts a full mesh backplane between disaggregated compute components within a rack and employs dedicated interfaces on each point-to-point link is wasteful and expensive. In this paper, we propose and describe two (i.e., electrical, and electrical-optical) variants of a network for composable DC (NetCoD). NetCoD adopts a targeted design to reduce the number of transceivers required when a mesh physical backplane is deployed between disaggregated compute components in the same rack. The targeted design leverages optical communication techniques and components to achieve this with minimal or no network performance degradation. We formulate a MILP model to evaluate the performance of both variants of NetCoD in rack-scale composable DCs that implement different forms of disaggregation. The electrical-optical variant of NetCoD achieves similar performance as a reference network while utilizing fewer transceivers per compute node. The targeted adoption of optical technologies by both variants of NetCoD achieves greater (4 - 5 times greater) utilization of available network throughput than the reference network which implements a generic design. Under the various forms of disaggregation considered, both variant of NetCoD achieve near-optimal compute energy efficiency in the composable DC while satisfying both compute and network constraints. This is because marginal concession of optimal compute energy efficiency is often required to achieve overall optimal energy efficiency in composable DCs.**

## I. Introduction

Data centers (DCs) are pivotal infrastructures which support on-demand access to computing capacity at scale. To meet present and future demands for on-demand computation services, there is a proliferation of the number of DC deployments on a global scale. Over the years, several efforts have been made to improve the efficiency of DCs to reduce capital expenditure (CAPEX) and operational expenditure (OPEX) and to improve their eco-friendliness. In recent times, the composable DC paradigm, which promotes dynamic orchestration of disaggregated computing components over suitable networks, is widely acknowledged as a tool for achieving further improvements in the efficiency of next generation DCs.

In the last two decades, several efforts have been made to improve the energy efficiency of core, metro and access communication networks that collectively support on-demand access to remote computing capacities in cloud and fog DCs [1]–[10]. Furthermore, several studies have also been conducted to improve traditional DC networks (DCNs) [11]–[13]. However, the advent of composable DCs demands significant revisions in DCN design. Designing a suitable network for composable DCs is daunting since disaggregation of traditional server resource components implies that high-bandwidth and ultra-low latency inter-resource communication must traverse higher tiers of DCNs. A range of switch architectures, network types and physical network topologies have been proposed for composable DCs in the literature. Section II of this paper briefly reviews various network topologies proposed for composable DCs.

The mesh physical topology is often proposed [14]–[16] to interconnect disaggregated components within a rack, because of the high capacity and low latency offered by this topology. Although, a full mesh physical topology is desirable to interconnect disaggregated components within the rack of a composable DC, adopting a generic design that requires dedicated transceivers on each point-to-point link of the mesh fabric is an overkill as the cost may out-weigh the benefits. Furthermore, inappropriate use of electrical switches in composable DCs can lead to significant increase in total DC power consumption [17]. In this paper, we leverage optical components, optical networking techniques, and technologies to minimize the number of transceivers per node while maintaining full mesh connectivity within each rack in the composable DC. We propose and describe two (i.e., electrical, and electrical-optical) variants of a network topology for rack-scale composable DCs. The network adopts a more targeted design while maintaining the full mesh physical topology

This work was supported by the Engineering and Physical Sciences Research Council (EPSRC), in part by INTelligent Energy aware NETworks (INTERNET) under Grant EP/H040536/1, in part by SwiTching And tRansmission (STAR) under Grant EP/K016873/1, and in part by Terabit Bidirectional Multi-user Optical Wireless System (TOWS) project under Grant EP/S016570/1. All data is provided in the results section of this paper. The first author would like to acknowledge his PhD scholarship awarded by the Petroleum Technology Trust Fund (PTDF), Nigeria.

The authors are with the School of Electronic and Electrical Engineering, University of Leeds, Leeds, LS2 9JT, U.K. (e-mail: el14oa@leeds.ac.uk; t.e.h.elgorashi@leeds.ac.uk; j.m.h.elmirghani@leeds.ac.uk).

within each rack of a composable DC. Relative to the generic design adopted in the literature when a mesh topology is adopted in the intra-rack backplane of composable DCs, the targeted design adopted in our proposed network requires fewer interfaces per compute node. We demonstrate the efficacy of both variants of the proposed network via mixed integer linear programming (MILP) optimization model formulation. By solving the MILP model under different scenarios, we show that the targeted design achieves the expected performance. Hence, network design challenges are mitigated, and cost is minimized by optimally utilizing the available network throughput. At the same time, the inherent waste, which is associated with the adoption of a generic design, is prevented. Furthermore, the electrical-optical variant of the proposed topology strategically utilizes electrical switches to minimize network power consumption and to maximize network utilization. This paper extends our initial work in [18] in the following ways.

- A brief review of network topologies proposed for composable DCs is made.
- Use of semiconductor-optical-amplifiers (SOA) based optical switches in a new configuration eliminate the need for optical filters in the proposed network topology.
- Electrical and electrical-optical variants of the proposed network topology are described.
- A complete MILP model is given for the first time.
- In addition to logical disaggregation, physical and logical disaggregation are considered at rack-scale.
- Lastly, results are discussed extensively.

The remainder of this paper is organized as follows: in Section II, an overview of composable DCs is given by briefly reviewing resource disaggregation and suitable network topologies for composable DC. Section III gives a description of the electrical and electrical-optical variants of the proposed network for composable DCs. Section IV presents the MILP models formulated to represent all network topologies studied in this paper and evaluates the performance of each topology by conducting a maximum throughput test and energy efficient network load test. Section V introduces the MILP model for energy efficient placement of VMs in composable DCs that adopt the proposed topology while Section VI presents results obtained by solving the MILP model in a rack that implements physical, logical and hybrid disaggregation. Finally, this paper is concluded in Section VII.

## II. Review of Composable DC Infrastructure

A composable DC comprises of physically and/or logically disaggregated computing components. These disaggregated components are composed, de-composed and re-composed on-demand via software over suitable networks to provision right-sized logical servers. The logical servers provide temporal support for applications. Hence, a dynamic DC, which achieves more granular and modular utilization of DC resources relative to today's server centric DCs, is enabled. Consequently, greater agility, flexibility and improved efficiencies are made possible in composable DCs.

### *A. Resource Disaggregation*

Resource disaggregation mitigates the resource stranding problem associated with the server centric DCs to enable greater utilization of resource components in cloud DCs [17] and fog DCs [19]. This is achieved by physically and/or logically separating DC computing components into homogenous or heterogenous resourced pools. Furthermore, utilization scopes may also be enforced between disaggregated components to ensure that application specific requirements are satisfied when provisioning logical servers in the composable DC. This is because some applications may require/desire specific forms of disaggregation to achieved optimal performance.

**Physical disaggregation:** Under this form of disaggregation, computing components are physically separated into homogenous resourced pools. A homogenous resourced pool is a server-like node which comprises of computing components of the same type (i.e., CPU, memory, or storage). Such homogenous nodes are subsequently allocated to racks in the composable DC at different scales i.e., rack-scale, pod-scale, or DC-scale [17], [20], [21]. At rack-scale, the DC comprises of many racks; each rack comprises of many homogenous resourced nodes of different resource types; and the resources within each node can only be used in conjunction with resources of other co-rack nodes to form a logical server. At pod-scale, the DC comprises of many racks; each rack holds multiple homogenous resourced nodes of the same resource type; heterogeneous resourced racks are allocated to each pod; and the resources within each node can only be used in conjunction with resources of other co-pod nodes to form a logical server. At DC-scale, the DC comprises of many homogenous resourced pods of different resource type; and the resources within each pod can be used in conjunction with resources of other pods in the DC to form a logical server.

**Logical disaggregation**: The utilization scope of logically disaggregated computing components is not enforced physically as observed for physically disaggregated DCs. Rather, the utilization is enforced virtually on-demand using knowledge of the infrastructure state and of application specific demands. Logical disaggregation supports re-purposing of the server-centric architecture of traditional DCs to enable a rack-scale composable DC that can support all type of applications [17]. Furthermore, logical disaggregation can also be implemented for any scale of physical disaggregation.

**Hybrid disaggregation**: A DC that implements hybrid disaggregation combines both physical and logical disaggregation to achieve optimal efficiency with zero or minimal violation of application specific requirements. Some compute nodes allocated to racks and pod in a DC that implements hybrid disaggregation are homogenously resourced while other are heterogenously resourced nodes (like a traditional servers).

### *B. Network Topologies for Composable DCs*

High capacity and ultra-low latency network topologies are required to interconnect nodes in a composable DC following physical, logical and hybrid disaggregation of computing components. Such networks support the orchestration and management software of composable DCs to optimally utilize disaggregated components. A range of network topologies have been proposed for composable DCs in the literature. Such topologies can be categorized as using switch architecture, network type, convergence, and physical topology as classification metrics.

*1) Switch Architecture*

Network topologies for composable DCs can also be classified into centralized and distributed switch networks based on the switch architecture. A network topology that solely performs switching and forwarding at centralized switches is a centralized switch architecture. The topology proposed in [22] and [23] implements a centralized switch architecture. On the other hand, a network topology that performs switching and forwarding functions at compute nodes either solely or to complement centralized switching and forwarding functions is a distributed switching architecture. The topologies proposed for composable DCs in [14], [16] and [15] implement a distributed switch architecture. A distributed switch architecture is more intelligent and adaptive realtive to a centralized switch. However, a disributed switch is relatively more complex and costly.

*2) Networks Type*

Classification based on network type considers the type of switching and forwarding components adopted in the corresponding network topology. Hence, topologies can be categorized as electrical, optical and hybrid network topologies. It is important to note that the use of optical links in composable DCs has no impact on this classification. This is because of the wide adoption of optical links in modern DCs.

- **Electrical topologies:** The multi-tier Ethernet based network topologies proposed for Intel's Rack-Scale Design (RSD) [23] reference model solely adopt electrical switches. Hence, it is an electrical network topology. Another multi-tier electrical topology is proposed for Huawei's high throughput computing DC architecture which is a rack-scale composable DC [14]. Similarly, the Gen-Z consortium proposed electrical switches for the implementation of a switched fabric to interconnect disaggregated CPU and memory modules [24]. Adoption of electrical switches can improve utilization of the optical fabrics that are often deployed in composable DCs. This is because electrical switches provide OEO conversions that can enable optimal utilization of optical links. Additionally, electrical topologies adopt a centralized switch architecture which is somewhat simpler and is well known due to its wide adoption in modern DCNs. However, electrical switches are known to have high power consumption [17]. Furthermore, compared to the high bandwidth communication required between disaggregated computing components in composable DCs, the capacity supported by traditional electrical switches are relatively lower. The latency of traditional electrical switches which ranges between 0.10 µs [16], and 10 µs [20] is unsuitable for seamless disaggregation of CPU and memory in composable DCs. Notwithstanding, the high bandwidth sub-100ns latency electrical switches proposed by the Gen-Z consortium are promising [25].
- **Optical topologies:** Optical topologies have been proposed for composable DCs to mitigate the challenges associated with electrical topologies. Therefore, the advantages of optical communication over electrical communication are leveraged. Such advantages include high-speed transparent communication, wavelength division multiplexing (WDM) and greater energy efficiency. The works in [15], [16], [26]–[28] have proposed various optical topologies for composable DCs. However, it is important to note that all-optical topologies also have inherent limitations. For example, all-optical topologies often require distributed switch architectures. This is because optical switches usually forward received wavelengths transparently in the absence of WDM conversation. Furthermore, established light-paths can be poorly utilized since multiplexing of traffic streams into lightpaths is performed at the source of the lightpath. This is because wavelength continuity must be maintained between source and destination nodes of a lightpath, and optical buffering capabilities are still limited.
- **Hybrid topologies:** Hybrid topologies for composable DCs leverage both electrical and optical switches. Hence, they may optimally utilize the benefits of both network types. An hybrid topology is proposed for rack-scale composable DC in [29]. Furthermore, the authors of [22] also proposed a hybrid topology for pod-scale composable DCs. However, it is important to note that poorly sited electrical switches in a hybrid network topology designed for composable DCs can be disadvantageous in terms of latency and power consumption [17]. A variant of the network for composable DCs proposed in this paper is a hybrid topology that demonstrates the strategic use of electrical and optical switches for optimal efficiency.

*3) Network Convergence*

Network topologies can be classified based on (non)separation of network traffic in composable DCs. Non-converged network topologies, as proposed in [29], have a dedicated fast fabric for high bandwidth and low-latency traffic types such a CPU-memory traffic while a second generic backplane supports other traffic types such as CPU-IO and CPU-disk traffic types. A non-converged network can enable reuse of traditional network infrastructure in a composable DC to implement the generic fabric. However, adoption of the composable DC paradigm is expected to lead to a complete overhaul of both compute and network infrastructures in DCs. On the other hand, a common fabric is adopted in a converged network topology. Hence, it requires a relatively simpler physical topology which can be optimally utilized. However, significantly higher network orchestration and control intelligence is required in a converged network. The works in [14]–[16], [23], [26] implement a converged netwotk topology.

*4) Physical Topology*

Network topologies proposed for composable DCs can also be classified based on the physical topology adopted in each design. Star, torus, and mesh physical topologies are commonly proposed for composable DCs networks. The authors of [14], [22], [24], [26] only proposed a star physical topology between compute nodes and switches within the same rack. Direct point-to-point connection is not employed between intra-rack compute nodes. A multi-tier star topology is another option proposed for in Intel's RSD [23]. The hubs (i.e., switches) in star physical topologies are a single point of failure, capacity bottlenecks and introduce additional delay due to single-hop communication. The authors of [14], [30] have also proposed the torus physical topology for composable DCs network topologies. Relative to the star topology, the torus topology requires higher number of links per node but provides higher path diversity and throughput. In addition to the star topology between the compute nodes and switches with the same rack, the authors of [14]–[16] also proposed full mesh connectivity between compute nodes in the same rack. The direct point-to-point optical connection between compute nodes enable low latency and high-capacity communicaton paths. Furthermore, a mesh topology provides greater path diversity relative to the torus topology. For example, the all-optical programmable disaggregated DCN (AOPD-DCN) [16] as illustrated in Fig. 1 proposes a mesh topology between nodes in the same rack while the optical ToR switch in each rack connects to an optical top of cluster (ToC) switch. The ToC switch in turn connects to an inter-cluster switch. An architecture on demand (AoD) optical switch capable of optical circuit and packet switching is employed to implement both ToC switch and the inter-cluster switch in the proposed DCN.

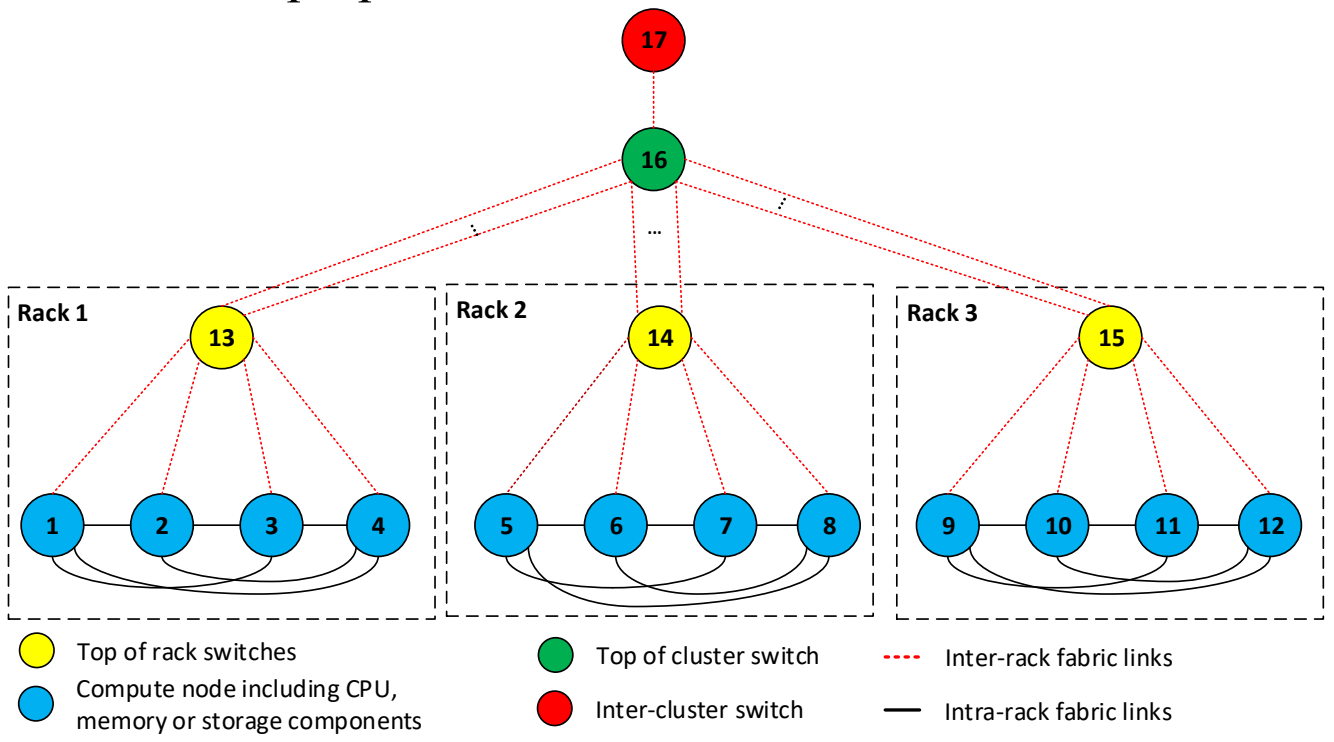

Fig. 1. All-optical programmable disaggregated DCN (AOPD-DCN) in a cluster of a composable DC.

However, the generic view that high data rate links are required concurrently between all co-rack compute nodes in a composable DC as proposed in [15], [16] can be very costly and wasteful. Although full-mesh physical connectivity is desirable for on-demand low latency and high-bandwidth communication between compute nodes, each compute node does not communicate concurrently with all co-rack nodes in a practical composable DC. For instance, consider a scenario where a traditional server with one CPU, one memory, one storage device and one network interface card (NIC) is physically disaggregated into four homogenous compute nodes. Given, CPU-to-memory traffic of 400 Gbps and 200 Gbps in the uplink and downlink directions respectively; CPU-to-storage traffic of 60 Gbps and 40 Gbps in the uplink and downlink directions respectively; and CPU-to-IO traffic of 10 Gbps and 8 Gbps in the uplink and downlink directions respectively, Equation (1) shows the corresponding traffic between the disaggregated compute components. The traffic shows that the compute node with CPU is a hotspot which also requires high-capacity interfaces. Similarly, the compute node with RAM also requires high-capacity interfaces because of the high-bandwidth communication with the remote CPU. Capacity requirement of compute nodes with hard disk drive (HDD) and NIC require low-medium interface capacity. Furthermore, in a rack with multiple disaggregated servers, it is unlikely that a compute node in the rack would communicate with all other co-rack compute nodes concurrently. Additionally, in a scenario where the compute node holds multiple CPU components, hence, a hotspot, it is unlikely in a practical scenario that such a node communicates with all other co-rack nodes at maximum capacity concurrently.

$$Traffic = \begin{bmatrix} \textbf{Node} & CPU & RAM & HDD & NIC \\ CPU & 0 & 400 & 60 & 10 \\ RAM & 200 & 0 & 0 & 0 \\ HDD & 40 & 0 & 0 & 0 \\ NIC & 8 & 0 & 0 & 0 \end{bmatrix} \quad (1)$$

In this paper, two variants of a practical Network for Composable DCs (NetCoD), which optimally utilize intra-rack physical mesh connectivity while using minimal number of interfaces, are proposed. The converged network topologies leverage optical communication techniques, components, and technologies for this purpose.

## III. Network Description

**Net**work for **Co**mposable **D**C (NetCoD) is a converged network topology that implements the distributed switch architecture. It leverages optical communication technologies and silicon photonics to support high-speed and low latency communication in composable DCs. Two variants of NetCoD are described in this paper i.e., electrical, and electrical-optical variants. Both variants of NetCoD are designed for a composable DC that implements resource disaggregation at rack-scale. Therefore, inter-resource communication is limited to the internal network of each rack while traditional DC traffic i.e., east-west, and north-south traffic traverse the inter-rack network of the DC. A common intra-rack network design is adopted in both variants of NetCoD. However, each variant integrates with a different inter-rack network type in composable DCs.

### *A. Intra-Rack Network*

The intra-rack network within each rack, as shown in Fig. 2, leverages optical communication components, technologies, and techniques to support high-speed and low latency communication between intra-rack resource components. Optical components such as optical backplane, optical circulators, combiners, demultiplexers, optical switches and silicon photonic transceivers are adopted at each compute node. The functions of the intra-rack network components are as follows:

**Passive optical backplane**: The optical backplane is a passive wavelength routing network within each rack that supports full mesh physical connectivity between nodes in the rack via point-to-point links. To minimize the size of the optical backplane within each rack, bi-directional transmission may be employed provided that the same wavelength is not active in forward and reverse directions simultaneously. Wavelength division multiplexing (WDM) enables increased transmission capacity over a point-to-point optical link between two compute nodes. Furthermore, because each optical link establishes a dedicated point-to-point communication link between unique node pairs in the rack, space division multiplexing (SDM) enables wavelength reuse on the optical backplane within the same rack.

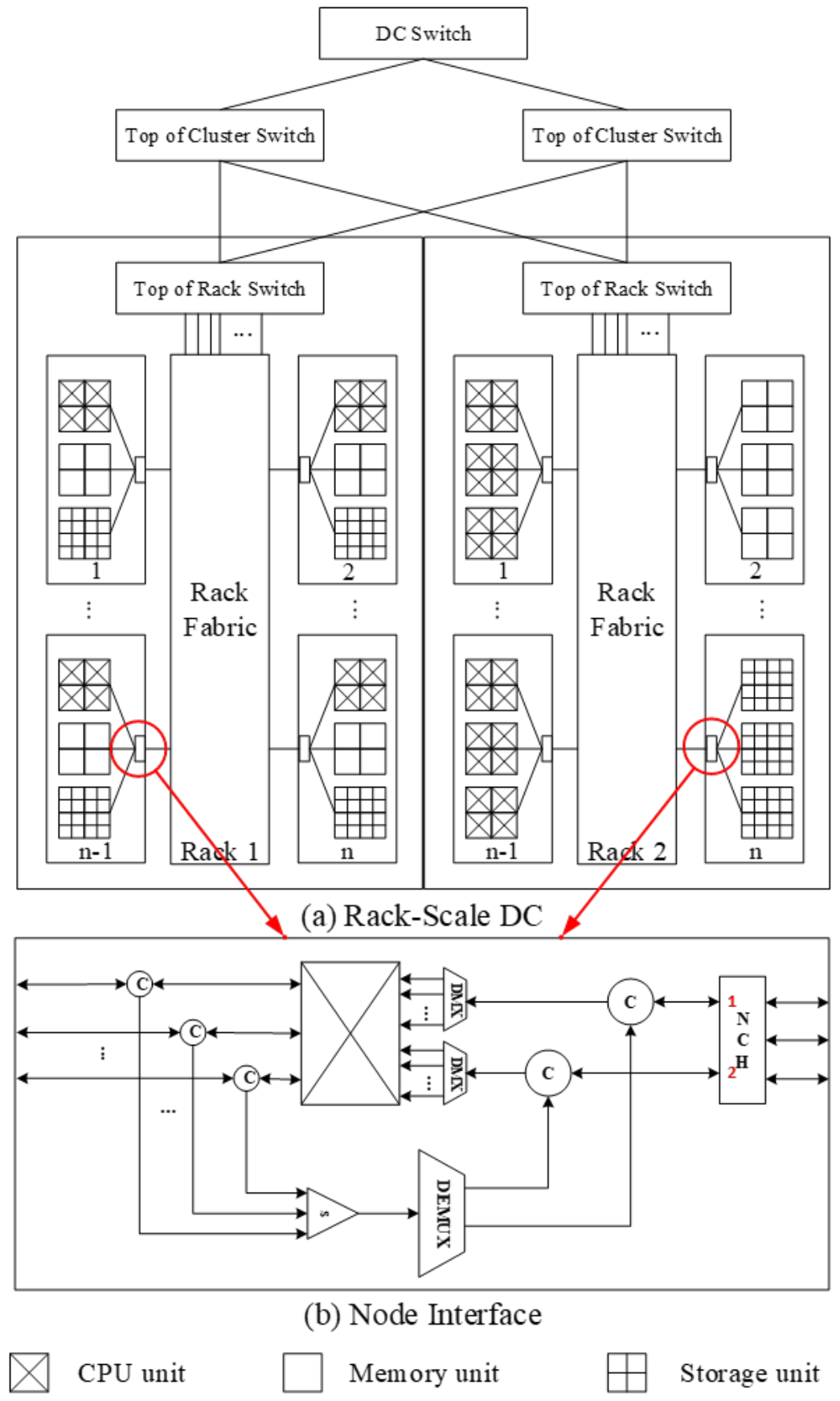

Fig. 2. Network for composable DCs (NetCoD)

**Node Controller Hub**: Each node in the rack-scale composable DC infrastructure has a node controller hub (NCH) which is proposed to replace the platform controller hub of traditional servers. As shown in Fig. 2, all compute components in a node are connected to the NCH. Compute components in the same node also maintain direct connectivity to one another via the node's on-board fabric to reduce the workload on the NCH, to ensure path diversity within the node and for greater energy efficiency. The NCH is a network element which performs network related computation in NetCoD. It may be implemented on an application-specific integrated circuit (ASIC) in commercial deployment and by a field-programmable gate array (FGPA) in experimental scenarios. The NCH performs the following functions:

- End-to-end virtual network setup (i.e., the routing function) for inter-nodal communication via direct or indirect physical links.
- Assignment of wavelengths for hop-to-hop communication (i.e., the forwarding function) over physical optical links.
- Multiplexing of data onto and the de-multiplexing of data from assigned inter-nodal wavelengths.
- Acting as an intermediate node on an indirect multi-hop path between two nodes.
- Optical switch path configuration to prevent wavelength collision on the passive optical backplane.
- Rate control and traffic scheduling as required to achieve optimal performance.

At each node, the NCH performs wavelength selection to avoid wavelength collision. Wavelength selection is performed based on global knowledge of the selection made at other nodes. Hence, all NCHs in NetCoD must be centrally controlled and orchestrated to ensure optimal wavelength utilization and the ability to operate NetCoD at maximum capacity.

**Integrated Interfaces**: Integrated with each compute node's NCH are two interfaces. Each interface comprises of an array of silicon photonic transceivers that transmit and receive a set of pre-defined wavelengths. The wavelengths transmitted by one interface are received by the other interface and vice versa. This enables a node to use all the wavelengths supported by its interface for transmission and reception of data concurrently. A common interface pair is deployed in all compute nodes within each rack to enable easy replication and to leverage the benefits of economies of scale. The interface setup at each node promotes wavelength reuse in each rack and minimizes the number of unique wavelengths required within each rack. Additionally, adoption of the interface pair at each node also enables path diversity which improves the resilience and capacity of NetCoD. The integration of a node's NCH element and the pair of interfaces may be implemented as a co-packaged device with optical IO by leveraging silicon photonics technologies.

**De-multiplexer**: In the transmitting direction, the de-multiplexers at each node separate the wavelengths transmitted from the interface into the appropriate port of the optical switch. On the other hand, in the receiving direction, the de-multiplexer receives multiplexed wavelengths directed to a corresponding node from the passive optical backplane and forwards each wavelength to the interface that should receive it. This is achieved via pre-configured physical connection between the de-multiplexer and the pair of interfaces attached to each node.

**Optical switches:** These are positioned before the point-to-point optical links between a node and the optical backplane to prevent wavelength collision on the optical backplane and at the receiving nodes. Path configuration on the optical switch should be performed by NCH based on global knowledge. An integrated and energy efficient SOA-based optical switch with low switching speed is proposed to implement the optical switch.

**Combiners:** In the receiving direction, the combiner at each compute node receives all wavelengths that have successfully traversed the optical backplane to reach the corresponding compute node and it combines and forwards the received wavelengths to the de-multiplexer.

**Optical Circulators:** Circulators enable bi-directional communication on optical links of the intra-rack backplane. Circulators are optional and may be employed between the optical backplane and the optical switch and between the de-multiplexers and the integrated interfaces of each compute node. Adoption of bi-directional communication can reduce the size of each rack's optical backplane by half relative to the use of unidirectional communication. However, use of bidirectional communication over an optical link in the optical backplane may limit the attainable capacity because wavelength utilization efficiency may reduce. Use of circulators to achieve bi-directional communication can also increase cost.

*1) Link Setup Process in Intra-Rack Network*

The following process is implemented to setup a link between two nodes within a rack that employs NetCoD. The NCH selects wavelengths (from the pool of wavelengths available at the pair of interfaces at each source node) which will ensure collision free transmission on the optical backplane and at the destination node. In the transmitting direction, the wavelengths transmitted by the interfaces of each node flow through optical circulators to the de-multiplexer, which separates all transmitted wavelengths of each node. The de-multiplexer is connected to an optical switch which directs the transmitted wavelengths to the appropriate link on the rack's optical backplane. Wavelength collision is avoided via the configuration of optical switches and via the use of parallel paths on the optical backplane to setup dedicated communication paths between each communicating nodes pair.

In the receiving direction, a combiner receives all transmitted wavelengths from other co-rack nodes and forwards the received wavelengths to a de-multiplexer. The de-multiplexer separates and forwards each received wavelength to the corresponding circulator that leads to the receiving interface. At the interface, each transceiver receives its associated wavelength and forwards the received data to the NCH. The NCH de-multiplexes the received data stream and forwards it to the appropriate compute component if it is in the destination node. Otherwise, the NCH forwards the received data to the corresponding interface linked to the next hop on the multi-hop communication path and selects an appropriate wavelength(s).

On the one hand, optical switches ensure that a wavelength is only transmitted to an intended destination node via the optical backplane. Combiners receive the ingress traffic (on the selected wavelengths destined for each node) from the optical backplane. Consequently, optical switches and combiners collectively reduce the number of interfaces required for each node to communicate over the full mesh optical backplane in a rack since concurrent all-to-all communication is not expected between all co-rack nodes.

As an illustration, consider a rack comprising of 4 compute nodes as illustrated in Fig 3, where interface 1 of each NCH emits wavelengths $\lambda0$ and $\lambda1$ and receives $\lambda2$ and $\lambda3$ while interface 2 of each NCH emits wavelengths $\lambda2$ and $\lambda3$ and receives wavelengths $\lambda0$ and $\lambda1$. Fig. 3 shows the wavelength assignment at each node that leads to maximum throughput in the intra-rack network without violating network constraints under unidirectional or bi-directional transmission mode on optical links. As illustrated in Fig. 3, Node 1 transmits wavelengths $\lambda0$-$\lambda3$ to Node 2; Node 2 transmits wavelengths $\lambda0$-$\lambda3$ to Node 4; Node 4 transmits wavelengths $\lambda0$-$\lambda3$ to Node 1; and Node 3 transmits wavelengths $\lambda0$-$\lambda3$ to Node 1. Hence, all nodes transmit and receive at full capacity by leveraging WDM. Furthermore, SDM enables wavelength reuse on disjoint physical links as shown in Fig. 3. It is important to note that the wavelength routing and assignment illustrated in Fig. 3 is a solution to a MILP model that maximizes throughput between four intra-rack nodes. Section IV gives a full description of the MILP model that was solved.

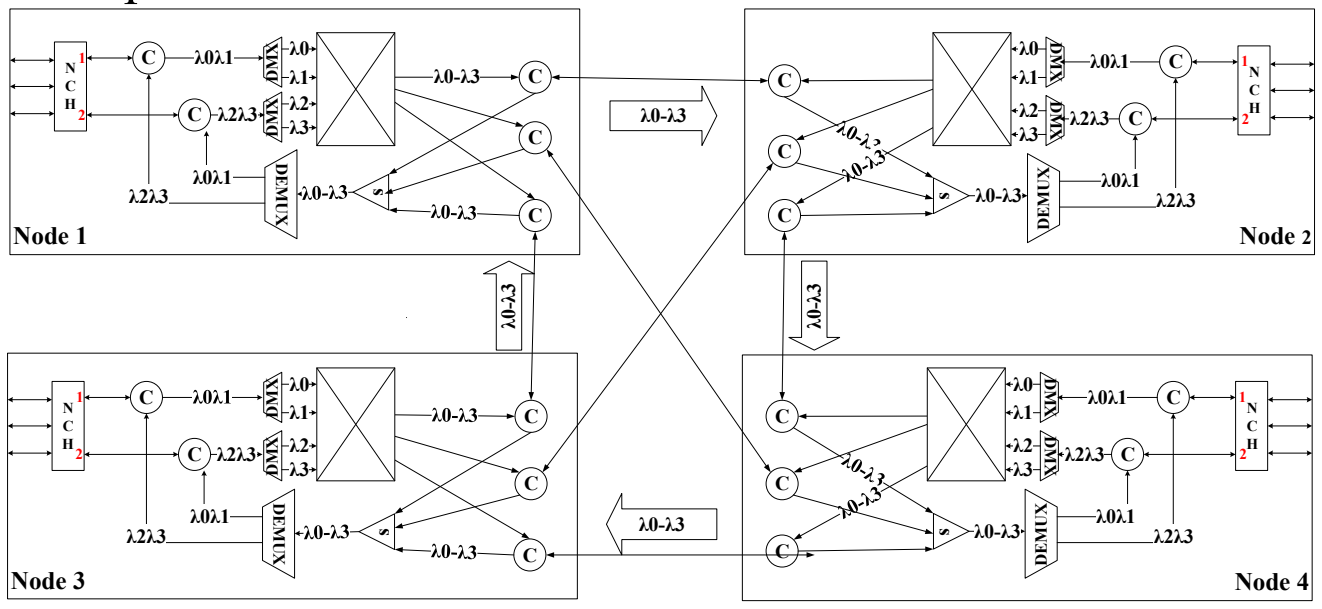

Fig. 3. Wavelength assignment between 4 intra-rack nodes in NetCoD.

## B. Inter-Rack Network

Two variants of the inter-rack network are proposed for NetCoD. The first variant, which is called electrical-NetCoD (E-NetCoD), adopts an electrical inter-rack network because it comprises of only electrical switches. The second variant, which is called electrical-optical-NetCoD (EO-NetCoD), adopts a hybrid inter-rack network which includes both electrical and optical switches. The physical topology depicted in Fig. 4 is adopted for both variants of NetCoD in the cluster of a composable DC.

*1) Electrical NetCoD*

In the electrical variant of NetCoD (E-NetCoD), the optical backplane of the intra-rack network also includes dedicated point-to-point links between compute nodes in each rack and a bespoke electrical leaf switch that functions as a top of rack (ToR) switch. The intra-rack network integrates with an electrical leaf-spine DCN topology via such links a shown in Fig. 4. The leaf switches are equipped with specialized interfaces to enable communication with compute nodes within the same rack via the NCH. The bespoke leaf switch within each rack and NCH (attached to each compute node in the rack) are centrally orchestrated to avoid wavelength collision. It is assumed that the bespoke leaf switch can perform wavelength conversion as required and that they have intrinsic intelligence to select wavelengths that avoid collision when communicating with each NCH. The leaf switch in each rack connects to the electrical spine (ToC) switches in the higher tier of the leaf–spine DCN topology and the spine (ToC) switches connect to electrical super-spine/gateway switch to support inter-cluster communication and north-south communication in the composable DC.

All electrical switches in the topologies performing routing and forwarding functions. However, the super-spine switch is expected to support higher capacity relative to other electrical switches in the leaf-spine physical topology. A leaf-spine network topology is employed in the inter-rack fabric because of its well-known advantages such as robustness enabled via path diversity and non-disruptive scalability. Additionally, the use of multiple aggregated physical links between switches in the inter-rack network may be implemented in large deployment scenarios, as shown in Fig. 4, to improve capacity as required. The integration of intra-rack network and inter-rack leaf-spine network of E-NetCoD conveys all traffic types (i.e., inter-resource traffic, east-west traffic, and north-south traffic) in the composable DC. Hence, a converged network. However, rack-scale disaggregation ensures that inter-resource traffic is limited to each rack; this prevents oversubscription and throughput challenges that may otherwise arise if such a converged network is deployed in a composable DC.

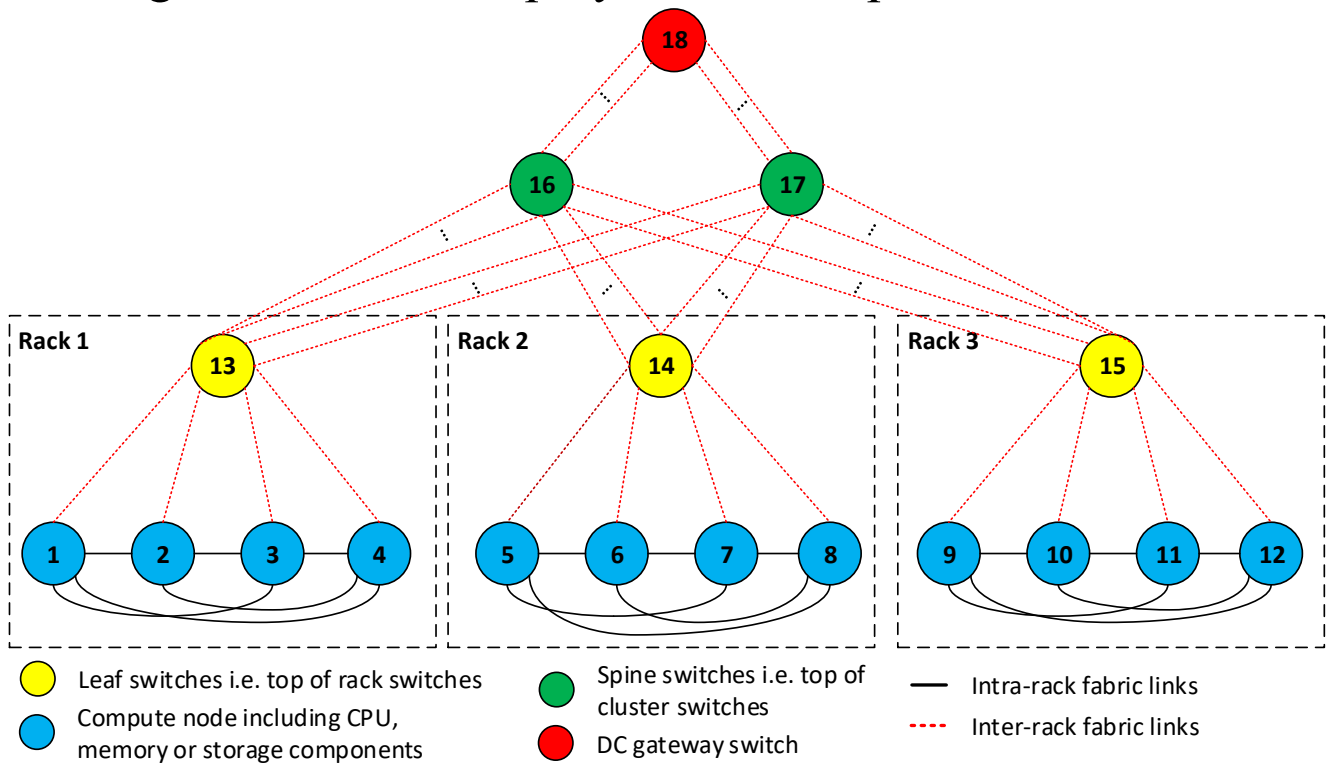

Fig. 4. An implementation of network for composable DCs in a single cluster.

In addition to supporting inter-rack traffic exchanges, the leaf switches can also function as an intermediate node for inter-resource traffic and east-west traffic exchange within the same rack. A low latency electrical switch such as the switch proposed by the Gen-Z consortium [24] may be adopted as the leaf switch. The integration of the optical backplane and the leaf-spine topology in each rack enables additional paths for inter-resource communication within the rack. Therefore, improving capacity and robustness. However, network bottlenecks resulting from the adoption of a shared medium for all communication types may occur at each node. It is expected that the higher capacity of single wavelength data rate in optical links will mitigate such bottlenecks. In recent times, up to 100 Gbps single wavelength transmission have been deployed [31] and even higher capacity is expected as optical technologies advance. Notwithstanding, a robust control mechanism is required to effectively manage the transmission of heterogeneous traffic types concurrently on the same media.

*2) Electrical-Optical NetCoD*

The electrical-optical variant of NetCoD (EO-NetCoD) adopts optical switches to replace electrical leaf and spine switches of the E-NetCoD while maintaining an electrical gateway switch. This reduces the OEO conversions in the network topology to enable reduced latency and reduced power consumption because each compute node can select appropriate wavelengths to establish both intra-rack and inter-rack light-paths. Alternatively, multiple light-paths may be established to facilitate inter-rack communication via intermediate nodes such as the electrical gateway switch or compute nodes in other racks of the composable DC. It is assumed that the high-capacity electrical gateway switch can perform wavelength conversion as required and that it also has intrinsic intelligence to select wavelengths that avoid collision when communicating over the inter-rack network.

A limitation of EO-NetCoD is the degradation in network performance resulting from wavelength continuity when routing is performed solely in the optical domain with limited wavelength and OEO conversions. Wavelength continuity leads to reduction in wavelength utilization and higher number of network connection request rejections in optical networks. Hence, it reduces network flexibility relative to a network that performs more OEO or wavelength conversions. However, this challenge may be mitigated when high single wavelength transmission rate is adopted. Since higher single wavelength transmission rate is constrained by technological advancement, the adoption of greater path diversity between switches in the inter-rack network is proposed to further mitigate the challenges introduced by wavelength continuity as given in Fig. 4. Factors that may determine the number of diverse paths provisioned between switches of the inter-rack network include but are not limited to the number of compute nodes in each rack, DC cluster size, size of the wavelength-pool supported in the NetCoD system and network availability criteria desired in the DC. It is expected that the adoption of rack-scale disaggregation will enable significant reductions in the volume of inter-rack traffic in the composable DC because only east-west traffic and north-south traffic types will traverse inter-rack network. Such traffic types have relatively lower bandwidth and higher tolerance to latency. Furthermore, a strategy that groups and places workloads with frequent east-west traffic exchange inside the same rack can also reduce the east-west traffic between racks in the composable DC.

In contrast to E-NetCoD where electrical switches enable wavelength and OEO conversions intrinsically, the adoption of optical switches in EO-NetCoD increases the likelihood of wavelength collision and consequently reduces wavelength reuse opportunities. On the one hand, the NCH element attached to each compute node can enable wavelength and OEO conversions by selecting appropriate wavelengths for hop-to-hop communication over both intra-rack and inter-rack networks. Furthermore, the electrical gateway switch forms an important boundary for wavelength collision and reuse in the inter-rack network of EO-NetCoD. To complement similar functions performed by the NCH, the gateway switch also performs OEO and wavelength conversions. The boundary introduced by the gateway switch limits the wavelength collision domain to each cluster of the composable DC. Hence, each cluster is an independent wavelength collision domain, and the pool of supported wavelengths can be independently reused in each cluster of the composable DC to maximize wavelength utilization and the total network capacity. Additionally, the capacity of links between optical switches in the inter-rack network is limited because wavelength collision avoidance is required in the all-optical layer. To overcome this limitation, path diversity should be employed between switches in the inter-rack network to improve capacity in large deployment scenarios.

It is important to note that a passive all-optical switch such as the arrayed waveguide grating router (AWGR) may be used to implement the leaf-spine layer of the EO-NetCoD. However, there are inherent disadvantages of using a passive optical switch which has a fixed routing matrix. Such design can reduce cost efficiency and energy efficiency because multiple transceivers must be fitted onto each node's interface. For example, if a single rack with 48 servers is considered, to achieve full mesh connectivity between all servers in that rack via single hop communication through an AWGR (without the use of time slots on wavelengths), a 48×48 AWGR is required, and each nodes interface must support the transmission and reception of 48 unique wavelengths. On the other hand, the use of multi-hop communication path to achieve virtual full mesh connectivity implies that routing and forwarding costs (power consumption) must be incurred at intermediate nodes on the communication path. In scenarios with high multi-hop communication, such power consumptions may outweigh any power savings achieved via the adoption of a passive optical switch with zero power consumption. Therefore, a configurable optical switch such as an optical cross connect (OXC) is proposed for EO-NetCoD to enable an adaptable and dynamic network for composable DCs.

*3) Scaling in NetCoD*

At the rack level, both variants of NetCoD scale-out via incremental and non-disruptive installation of additional compute nodes. The newly added compute nodes are connected to existing compute nodes and to the ToR switch in that rack via the passive optical backplane. At the cluster level, NetCoD scales-out via incremental and non-disruptive installation of more racks. The additional racks are connected to the dedicated leaf-spine inter-rack network of each cluster. Finally, at the DC-level, NetCoD supports on-demand scale-out via incremental and non-disruptive installation of more clusters which are connected to the electrical gateway switch of the composable DC.

## IV. MILP Model for Network Topologies

This section presents a MILP model that is formulated to optimize both variants of NetCoD. The MILP model is also revised to implement AOPD-DCN. The MILP model performs routing and forwarding of network traffic over the corresponding network topology to minimize or maximize a specific objective.

### *A. MILP Model for E-NetCoD*

The model sets, parameters, and variables for a composable DC that implements E-NetCoD are given as follows.

**Sets:**

| | |
|---|---|
| $CN$ | Set of compute nodes $CN \subseteq N$ |
| $DG$ | Set of DC gateway switches to the Internet $DG \subseteq N$ |
| $CG$ | Set of compute nodes and DC gateway switches $CG \subseteq N$; $CG = CN \cup DG$ |
| $A$ | Set of leaf and spine switches $A \subseteq N$ |
| $RFN$ | Set of routing and forwarding nodes ($RFN$) in the DC, $RFN \subseteq N$; $RFN = A \cup CN \cup DG$ |
| $N$ | Set of all nodes, $N = A \cup CN \cup DG$ |
| $N_m$ | Set of all neighbor nodes of node $m \in N$; $N_m \subseteq N$. |
| $IN_m$ | Set of all intra-rack neighbor nodes of node $m \in N$; $IN_m \subseteq N$. |
| $CN_m$ | Set of all compute nodes that are neighbors of compute node $m \in CN$; $CN_m \subseteq CN$. |
| $W$ | Set of transmission wavelengths supported in the network. |
| $INT$ | Set of interfaces supported by a compute node. |

**Network Parameters:**

| | |
|---|---|
| $IT_{wf}$ | $IT_{wf} = 1$ if wavelength $w \in W$ is allocated to interface $f \in INT$ for transmission of data traffic, otherwise $IT_{wf} = 0$ |
| $IR_{wf}$ | $IR_{wf} = 1$ if wavelength $w \in W$ is allocated to interface $f \in INT$ for reception of data traffic, otherwise $IR_{wf} = 0$ |
| $\mu_i$ | Load proportional routing cost for a routing and forwarding node $i \in RFN$, (J/b) |
| $OBepb$ | On-board network interface energy per bit (J/b) |
| $TXepb_m$ | Transmitting energy per bit (J/b) of routing and forwarding node $m \in RFN$ |
| $RXepb_m$ | Receiving energy per bit (J/b) of routing and forwarding node $m \in RFN$ |
| $RLepb_m$ | Relaying energy per bit (J/b) of routing and forwarding node $m \in RFN$ |
| $OXP$ | Optical switch operational power in Watt |
| $ESIPC$ | Electrical switch operational power in Watt |
| $SOASWepb$ | SOA switch energy per bit (J/b) |
| $\rho$ | Number of spine switches in the composable DC |
| $\alpha$ | Number of electrical gateway or super-spine switches in the composable DC |
| $NAR$ | Number of active racks in the composable DC |
| $\tau_{sd}$ | Total traffic from node $s \in RFN$ to node $d \in RFN$. |
| $D$ | Maximum data rate of a single wavelength. |
| $Q$ | A big number (eg. 100000) |
| $G$ | A big number (eg. 1000) |

**Network Variables:**

| | |
|---|---|
| $T_{sd}$ | Total traffic from node $s \in RFN$ to node $d \in RFN$. |
| $C_{sd}^{ij}$ | Volume of $T_{sd}$ traversing virtual link $(i, j)$. $i \in RFN, j \in RFN, s \in RFN, d \in$ |
| | $RFN: i \neq j, s \neq d$. It denotes routing of traffic in the virtual network. |
| $C_{ij}$ | Volume of traffic on virtual link $(i,j)$; $i \in RFN, j \in RFN$ |
| $\Phi_i$ | Traffic transmitted at routing node $i \in RFN$ |
| $\Psi_i$ | Traffic received at routing node $i \in RFN$ |
| $\Omega_i$ | Traffic relayed at routing node $i \in RFN$ |
| $\phi_m$ | Traffic transmitted at forwarding node $m \in RFN$ in the physical layer. |
| $\psi_m$ | Traffic received at forwarding node $m \in RFN$ in the physical layer. |
| $\omega_m$ | Traffic relayed at forwarding node $m \in RFN$ in the physical layer. |
| $W_{wmn}^{ij}$ | Volume of traffic on virtual link $(i,j)$ that uses wavelength $w \in W$ on physical link $(m,n), i \in RFN, j \in RFN, m \in N, n \in N_m: i \neq j, m \neq n$ |
| $W_{wmn}$ | Volume of traffic that uses wavelength $w \in W$ on physical link $(m,n)$, $m \in N, n \in N_m: m \neq n$ |
| $F_{wmn}$ | $F_{wmn} = 1$ if $W_{wmn} > 0$. Otherwise, $F_{wmn} = 0, w \in W, m \in N, n \in IN_m: m \neq n$ |
| $\mathcal{F}_{wmn}$ | $\mathcal{F}_{wmn} = 1$, if $F_{wmn} \vee F_{wnm} = 1$. Otherwise, $\mathcal{F}_{wmn} = 0, w \in W, m \in N, n \in IN_m: m \neq n$ |
| $Y_{wfm}$ | $Y_{wfm} = 1$ if wavelength $w \in W$ is used on interface $f \in INT$ of compute node $m \in CN$ to transmit traffic to neighbor nodes or receive traffic from neighbor nodes. Otherwise, $Y_{wfm} = 0$. |

The variables are related as follows.

$$W_{wmn} \geq F_{wmn} \quad (2)$$
$$\forall\, w \in W, \forall\, m \in N, n \in IN_m: m \neq n$$

$$W_{wmn} \leq Q\, F_{wmn} \quad (3)$$
$$\forall\, w \in W, \forall\, m \in N, n \in IN_m: m \neq n$$

Equations (2) and (3) derive the state of each wavelength available on each physical link within the rack.

$$\Phi_i = \sum_{j \in RFN} \sum_{d \in RFN} C_{id}^{ij} \quad (4)$$
$$\forall i \in RFN: d \neq i, i \neq j$$

Equation (4) derives the traffic transmitted by a routing and forwarding node in the virtual layer of the network topology.

$$\Psi_i = \sum_{s \in RFN} \sum_{j \in RFN} C_{si}^{ji} \quad (5)$$
$$\forall i \in RFN: s \neq i, i \neq j$$

Equation (5) derives the traffic received by a routing and forwarding node in the virtual layer of the network topology.

$$\Omega_i = \sum_{s \in RFN} \sum_{d \in RFN} \sum_{j \in RFN} C_{sd}^{ij} \quad (6)$$
$$\forall i \in RFN: s \neq d, s \neq i, d \neq i, i \neq j$$

Equation (6) derives the traffic relayed by a routing and forwarding node in the virtual layer of the network topology.

$$\phi_m = \sum_{w \in W} \sum_{n \in N_m} \sum_{j \in RFN} W_{wmn}^{mj} \quad (7)$$
$$\forall m \in RFN: j \neq m, m \neq n$$

Equation (7) derives the traffic transmitted by a routing and forwarding node in the physical layer of the network topology.

$$\psi_m = \sum_{w \in W} \sum_{n \in N_m} \sum_{i \in RFN} W_{wnm}^{im} \quad (8)$$
$$\forall m \in RFN: i \neq m, m \neq n$$

Equation (8) derives the traffic received by a routing and forwarding node in the physical layer of the network topology.

$$\omega_m = \sum_{w \in W} \sum_{n \in N_m} \sum_{j \in RFN} \sum_{i \in RFN} W_{wmn}^{ij} \quad (9)$$
$$\forall m \in RFN: i \neq j, i \neq m, j \neq m, m \neq n$$

Equation (9) derives the traffic relayed by a routing and forwarding node in the physical layer of the network topology.

$$TNRP = \mu_i \sum_{i \in RFN} (\Phi_i + \Psi_i + \Omega_i) \quad (10)$$

Equation (10) derives the total network routing power ($TNRP$) due to the routing function performed by routing and forwarding nodes in the virtual layer. This represents any additional power consumed by nodes that can perform routing in the virtual layer.

$$TNFP = \sum_{i \in RFN} (\phi_i\, TXepb_i + \psi_i\, RXepb_i + \omega_i\, RLepb_i) \quad (11)$$

Equation (11) determines the total network forwarding power ($TNFP$) due to the forwarding function performed by routing and forwarding nodes in the physical layer.

$$TXNP = \sum_{i \in CN} (\phi_i + \omega_i)\, SOAepb + (NAR + \rho + \alpha) ESIPC \quad (12)$$

Equation (12) establishes the total other network power ($TXNP$) for E-NetCoD which is measured by the power consumed by active physical node/components in the network topology. It comprises of the power consumed by SOA switches at compute nodes and the fixed operational power of electrical switches in the composable DC.

**Total Network Power Consumption**

$$TNPC = TNFP + TNRP + TXNP \quad (13)$$

The total network power consumption ($TNPC$) is a sum of the $TNFP$, $TNRP$ and $TXNP$.

The model MILP is defined as follows:

Two objective functions are considered for the model to represent two distinct scenarios.

**Objective 1**: Maximize:

$$\sum_{s \in CG} \sum_{d \in CG: s \neq d} T_{sd} \quad (14)$$

Equation (14) is the first objective function that maximizes the total throughput between all communicating nodes pairs in the composable DC. This objective function maximizes the total traffic exchanged between selected routing and forwarding nodes in the DC.

**Objective 2**: Minimize:

$$TNPC \quad (15)$$

Equation (15) is the second objective function. It minimizes the total network power consumed by routing and forwarding the input traffic $\tau_{sd}$ over the corresponding network topology under consideration.

**Subject to:**

$$\sum_{j\in RFN:i\neq j} C_{sd}^{ij} - \sum_{j\in RFN:i\neq j} C_{sd}^{ji} = \begin{cases} T_{sd} & i=s \\ -T_{sd} & i=d \\ 0 & otherwise \end{cases} \quad (16)$$

$$\forall i\in RFN, \forall s,d\in CG: s\neq d$$

Constraint (16) enforces flow conservation in the virtual layer setup between electronic routing and forwarding nodes in the composable DC. Note that for objective 2, $T_{sd}$ should be replaced by $\tau_{sd}$ in (16).

$$\sum_{s\in CG}\sum_{d\in CG:s\neq d} C_{sd}^{ij} = C_{ij} \quad (17)$$

$$\forall i\in RFN, \forall j\in RFN: i\neq j$$

Constraint (17) calculates the volume of traffic on each virtual link provisioned between a pair of routing and forwarding nodes in the virtual layer.

$$\sum_{w\in W}\sum_{n\in N_m:m\neq n} W_{wmn}^{ij} - \sum_{w\in W}\sum_{n\in N_m:m\neq n} W_{wnm}^{ij} = \begin{cases} C_{ij} & m=i \\ -C_{ij} & m=j \\ 0 & otherwise \end{cases} \quad (18)$$

$$\forall m\in N, \forall i,j\in RFN: i\neq j$$

Constraint (18) enforces flow conservation in the physical network topology between all nodes in the DC.

$$\sum_{i\in RFN}\sum_{j\in RFN:i\neq j} W_{wmn}^{ij} = W_{wmn} \quad (19)$$

$$\forall w\in W, \forall m\in N, \forall n\in N_m: m\neq n$$

Constraint (19) calculates the volume of traffic on each wavelength on a physical link in the network topology.

$$W_{wmn} \leq D \quad (20)$$

$$\forall w\in W, \forall m\in N, n\in N_m: m\neq n$$

Constraint (20) is the capacity constraint of each wavelength that is used on a physical link.

$$\sum_{n\in IN_m}\sum_{f\in INT} F_{wmn}\, IT_{fw} \leq 1 \quad (21)$$

$$\forall w\in W, \forall m\in CN: m\neq n$$

$$\sum_{n\in IN_m}\sum_{f\in INT} F_{wnm}\, IR_{fw} \leq 1 \quad (22)$$

$$\forall w\in W, \forall m\in CN: m\neq n$$

Constraint (21) ensures that each wavelength transmitted by a compute node is transmitted once from that node by an interface that is designed to emit that wavelength. On the other hand, Constraint (22) ensures that each wavelength received by a compute node is received once at an interface that is designed to receive that wavelength.

$$\sum_{n\in IN_m:m\neq n} F_{wmn}\, IT_{fw} + \sum_{n\in IN_m:m\neq n} F_{wnm}\, IR_{fw} = Y_{wfm} \quad (23)$$

$$\forall w\in W, \forall f\in INT, \forall m\in CN$$

Constraint (23) ensures that the same wavelength does not flow in opposite directions at a given interface of a compute node. This is required when bi-directional communication is employed on the physical link that connects each interface to the optical backplane and each interface comprises of an array of unique transceivers.

$$F_{wmn} + F_{wnm} = \mathcal{F}_{wmn} \quad (24)$$

$$\forall w\in W, \forall m\in CN, n\in IN_m: m\neq n$$

Constraint (24) ensures that same wavelength does not flow in opposite directions in a physical link on each rack's optical backplane. This constraint implements bi-directional communication on the passive intra-rack backplane. It is not active when unidirectional communication is implemented on the passive intra-rack backplane.

### *B. MILP Model for EO-NetCoD*

In contrast to E-NetCoD, some network constraints must be revised to represent EO-NetCoD in a MILP model while others must be introduced. Consequently, additional set, parameter and variables are introduced while others are revised. The additional set enable the representation of revised nodes when EO-NetCoD is implemented. The hybrid inter-rack network is created via the adoption of optical switches to replace electrical leaf and spine switches. It comprises of all nodes (compute nodes, optical switches, and DC gateway switch) which are connected directly to any optical switch in the physical network topology. The additional variables enable representation of traffic routing over EO-NetCoD.

**Revised and additional Sets and Parameter**

| | |
|---|---|
| $RFN$ | Set of all routing and forwarding nodes $RFN = CG = CN \cup DG$ |
| $N$ | Set of all nodes $N = OS \cup CN \cup DG$ |
| $N_m$ | Set of all neighbor nodes of node $m\in N, N_m \subseteq N$ |
| $OS$ | Set of optical switches, $OS \subseteq N$ |
| $FN_m$ | Set of all neighbor nodes of node $m\in N$; $FN_m \subseteq N$ which are part of the hybrid inter-rack network. |
| $\kappa$ | Cost associated with each path provisioned in an optical switch in Watts |

**Network Variables:**

| | |
|---|---|
| $U_{ij}$ | Volume of traffic on virtual link $(i,j)$, $i\in RFN, j\in RFN$, that traverses intra-rack network. |
| $V_{wij}$ | Volume of traffic using wavelength $w\in W$ on virtual link $(i,j)$, $i\in RFN, j\in RFN$, that traverses the hybrid inter-rack network. |
| $H_{wij}$ | $H_{wij}=1$ if $V_{wij}>0$. Otherwise $H_{wij}=0, w\in W, i\in RFN, j\in RFN: i\neq j$ |
| $X_{wman}$ | The configured switching matrix of an optical switch. $X_{wman}=1$ if the wavelength $w\in W$ from node $m\in FN_a$ enters optical switch $a\in OS$ and is relayed to node $n\in FN_a$. Otherwise, $X_{wman}=0$. |
| $B_{wman}^{ij}$ | $B_{wman}^{ij}$ gives the traffic $W_{wma}^{ij}$ that enters optical switch $a\in OS: a\in FN_m$ from node $m\in FN_a$ and is relayed to node $n\in FN_a$ on the hybrid inter-rack network. $w\in W, m\in N$ |

The EO-NetCoD comprises of two intrinsic networks i.e., an intra-rack network between routing and forwarding capable compute nodes within the same rack and a hybrid inter-rack network enabled by the deployment of optical switches in a leaf-spine topology. Point-to-point light-paths are setup between routing and forwarding nodes over the physical links of the hybrid inter-rack network. The passive nature of the optical switches after path configuration implies that the optical switches are only aware of directly connected neighbors on the hybrid inter-rack network.

The $TXNP$ for EO-NetCoD can be measured by the power consumed by active physical node/components in the network topology. It comprises of the power consumed by SOA switches at compute nodes, the fixed operating power of optical and electrical switches and the cost of setting up optical paths in each optical switch in the composable DC. It is derived as by Equation (25).

$$TXNP = \sum_{i \in CN} (\phi_i + \omega_i)\, SOAepb + (NAR + \rho)\, OXP + \alpha\, ESIPC + \kappa \left( \sum_{w \in W} \sum_{a \in OS} \sum_{m \in FN_a} \sum_{n \in FN_a} X_{wamn} \right) \quad (25)$$

Given the traffic demand ($T_{sd}$ or $\tau_{sd}$) in the DC, virtual links $C_{ij}$ are setup between routing and forwarding nodes in the DC to efficiently route traffic over the network topology as seen earlier in Constraints (16) and (17). However, a virtual link $C_{ij}$ between two routing and forwarding nodes in EO-NetCoD can traverse the intra-rack network and/or the hybrid inter-rack network. As a result, Constraint (18) is no longer applicable in such a setup. To accommodate such a unique setup in the MILP model, the following network constraints are introduced.

$$U_{ij} + \sum_{w \in W} V_{wij} = C_{ij} \quad (26)$$

$$\forall\, i \in RFN, \forall\, j \in RFN: i \neq j$$

Constraint (26) ensures that the volume of traffic on virtual link $C_{ij}$ is equal to the sum of traffic sent via the intra-rack network and the hybrid inter-rack network. This is because the virtual link can be routed via the intra-rack network or/and via the inter-rack network.

$$\sum_{w \in W} \sum_{n \in CN_m: m \neq n} W_{wmn}^{ij} - \sum_{w \in W} \sum_{j \in CN_m: m \neq n} W_{wnm}^{ij} = \begin{cases} U_{ij} & m = i \\ -U_{ij} & m = j \\ 0 & otherwise \end{cases} \quad (27)$$

$$\forall\, i,j \in RFN: s \neq d\ \forall\, m \in CN$$

Constraint (27) enforces flow conservation in physical links of the intra-rack network of each rack in the DC.

$$\sum_{n \in FN_m: m \neq n} W_{wmn}^{ij} - \sum_{n \in FN_m: m \neq n} W_{wnm}^{ij} = \begin{cases} V_{wij} & m = i \\ -V_{wij} & m = j \\ 0 & otherwise \end{cases} \quad (28)$$

$$\forall\, m \in N, \forall\, i,j \in RFN: i \neq j, \forall\, w \in W$$

Constraint (28) enforces flow conservation in physical links of the hybrid inter-rack network in the DC. It also enforces wavelength continuity on each light-path created between two nodes in the inter-rack network.

$$V_{wij} \geq H_{wij} \quad (29)$$

$$\forall\, w \in W, \forall\, i \in RFN, j \in RFN: i \neq j$$

$$V_{wij} \leq Q\, H_{wij} \quad (30)$$

$$\forall\, w \in W, \forall\, i \in RFN, j \in RFN: i \neq j$$

Constraints (29) and (30) jointly derive the state of all potential light-paths that can traverse the hybrid inter-rack network.

$$\sum_{j \in RFN} H_{wij} \leq 1 \quad (31)$$

$$\forall\, w \in W, \forall\, i \in CN: i \neq j$$

Constraint (31) ensures that a wavelength is used to setup a single light-path from a compute node. Hence, the wavelength must not be used more than once on any light-path originating at this compute node. Note that the gateway switch is permitted to use a given wavelength on different light-paths provided that network routing constraints are not violated. This is because a sophisticated network switch is assumed. Multiple links emanate from the gateway switch i.e., path diversity is employed; hence, the risk of wavelength collision is mitigated since a wavelength can be re-used on disjoint links.

**Optical network routing constraints**

$$\sum_{n \in FN_a} X_{wman} \leq 1 \quad (32)$$

$$\forall\, w \in W, \forall\, a \in OS, m \in FN_a$$

$$\sum_{m \in FN_a} X_{wman} \leq 1 \quad (33)$$

$$\forall\, w \in W, \forall\, a \in OS, n \in FN_a$$

Constraint (32) ensures that an ingress wavelength to an optical switch from a given neighbor node of the optical switch is relayed to at most one neighbor node of that optical switch. On the other hand, Constraint (33) ensures that an egress wavelength from an optical switch, which is relayed to a neighbor node of the optical switch, entered the switch from only one neighbor node of the optical switch. This avoids wavelength collision at a given output port of the optical switch. Both constraints (32) and (33) implement a passive switching matrix for a configurable optical switch.

$$B_{wman}^{ij} \leq W_{wma}^{ij} \quad (34)$$

$$\forall\, w \in W, \forall\, i,j \in RFN, \forall\, a \in OS, \forall\, m,n \in FN_a$$

$$B_{wman}^{ij} \leq D\, X_{wman} \quad (35)$$

$$\forall\, w \in W, \forall\, i,j \in RFN, \forall\, a \in OS, \forall\, m,n \in FN_a$$

$$B_{wman}^{ij} \geq W_{wma}^{ij} - D(1 - X_{wman}) \quad (36)$$

$$\forall\, w \in W, \forall\, i,j \in RFN, \forall\, a \in OS, \forall\, m,n \in FN_a$$

Constraints (34) - (36) linearize the derivation of continuous variable $B_{wman}^{ij}$ which includes a product of a continuous variable and a binary variable as shown in Constraint (37).

$$B_{wman}^{ij} = W_{wma}^{ij}\, X_{wman} \quad (37)$$

$$\forall\, w \in W, \forall\, i,j \in RFN, \forall\, a \in OS, \forall\, m,n \in FN_a$$

Constraint (37) ensures that the traffic that enters an optical switch is relayed according to the path configuration of that optical switch.

$$W_{wan}^{ij} = \sum_{m \in N: m \in FN_a} B_{wman}^{ij} \quad (38)$$

$$\forall w \in W, \forall i, j \in RFN, \forall a \in OS, \forall n \in FN_a$$

Constraint (38) allows optical switches to route traffic passively over the hybrid inter-rack network. This is achieved by ensuring that the traffic that enters an optical switch from a given neighbor node is relayed to the appropriate output port of the optical switch as specified by the configured switching matrix of the switch.

### C. MILP Model for AOPD-DCN

A similar approach taken to emulate optical switches in the MILP model for EO-NetCoD can be adopted to emulate the wavelength selective switch (WSS) deployed in AOPD-DCN. Additionally, to simplify model formulation we assume that the ToC (AoD) switches in AOPD-DCN are always configured to perform OCS while the inter-cluster AoD switch is setup to perform packet switching and forwarding. Since, no limitation is specified for interfaces and bi-directional communication is not considered, interface constraints (21) - (24) are not applicable in the intra-rack network of AOPD-DCN. In addition to other network constraints from Section IV.A and IV.B the following constraints are required to represent AOPD-DCN as MILP Model.

$$\sum_{n \in FN_m: m \neq n} F_{wmn} \leq 1 \quad (39)$$

$$\forall w \in W, \forall m \in CN$$

Constraint (39) ensures that a given wavelength from a compute node is transmitted only once by that compute node on the inter-rack network.

$$\sum_{n \in FN_m: m \neq n} F_{wnm} \leq 1 \quad (40)$$

$$\forall w \in W, \forall m \in CN$$

Constraint (40) ensures that a given wavelength received from any neighbor node of a given compute node on the inter-rack network is received only once at the compute node.

$$\sum_{w \in W} W_{wmn} \leq \mathbb{C} \quad (41)$$

$$\forall m \in CN, \forall n \in CN_m$$

Constraint (41) is the capacity constraint of the physical link between two compute nodes in the same rack. Where $\mathbb{C}$ is the maximum transmitting and receiving capacity supported on the point-to-point physical link between compute nodes on the intra-rack backplane.

### D. Network Setup and Input Parameter

We compare the performance of both variants of NetCoD to AOPD-DCN in the small cluster of a composable DC depicted in Fig. 1 and Fig. 4. The cluster comprises of 3 racks, each rack holds 4 compute nodes and a ToR switch. Each compute node is designed to perform network routing and forwarding functions along with wavelength selection. The ToR switches of both variants of NetCoD are connected to two spine switches at the top of the cluster. The spine switches subsequently connect to the gateway switch for inter-cluster and inter-DC communications. However, as shown in Fig. 1, physical connectivity is slightly different in the inter-rack network of AOPD-DCN. ToR switches connect to a single ToC switch that is configured to perform OCS connectivity between adjacent racks in the cluster. The ToC switch subsequently connects to inter-cluster switch, which provides inter-cluster communication in the composable DC, in the upper tier. The inter-cluster switch is setup to perform packet routing and forwarding as done by the DC gateway switch in both variants of NetCoD. In all evaluation scenario, the use of multiple links between two switches to achieve path diversity is not implemented or modelled to ensure simplicity. This assumption is practical in this evaluation scenario because a small cluster is considered. A composable DCs with more and bigger clusters would require the implementation of path diversity between network switches for robustness and to boost network capacity.

The load proportional energy per bit values of DCN tiers is given in Table I along with the operating power consumption of electrical and optical network components. We adopt the energy per bit values predicted for on-board, intra-rack and inter-DC tiers of next generation DCNs as given in [32]. However, because it is expected that the electrical ToR/ToC switch will be relatively more complex than the NCH but less complex than the DC gateway switch, the inter-rack energy per bit value suggested in [32] is not suitable in our setup. Hence, we conservatively assume that the energy per bit value for each electrical switch in the leaf-spine-layers of the inter-rack network is 5 pJ/b in all evaluation scenarios to reflect the relative difference in DCN tier complexity. The typical operating (idle) power of all electrical switches in the all network topologies consider is 312 W [33].

The NCH has two functions. Firstly, it has to convert the electrical data streams to an appropriate optical wavelength and transmit this data i.e., the forwarding function. The associated power consumption is typically 1 pJ/b [32]. Secondly, the NCH also has to compute the route to the destination and configure the SOA switches and set up the path. It is assumed here that these operations consume an equal amount of power in the NCH. Spreading this power consumption between all the data streams and wavelengths handled by the NCH leads to a power consumption of 1 pJ/b for path computation and setup. It should be noted that this choice is on the pessimistic side as the path computation tasks can consume much lower power if for example look up tables are used. Similarly, it is also conservatively assumed that path computation and setup functions performed by all electrical switches also leads to a power consumption of 1 pJ/b.

We adopt 100Gbps for single wavelength transmission in the network topology given recent practical demonstration of such lane rate [31]. We expect even greater single wavelength lane rate for short reach inter-connects in the future as optical technologies advance [34]. To simplify the evaluation scenario, it is assumed that each interface, which is integrated with the NCH of a compute node, can emit 4 distinct wavelengths. Hence, a maximum 8 wavelength is supported under both variants of NetCoD. For fair comparison, it is also assumed that the interface used by each compute node in AOPD-DCN to connect to the inter-rack network also supports 8 wavelengths. Hence, enabling 800 Gbps link from each compute node to the ToR switch. Additionally, it is assumed that each interface connected to the intra-rack backplane of AOPD-DCN can transmit or receive four wavelengths in parallel at 100 Gbps lane rate i.e., $\mathbb{C} = 400\ Gbps$. Hence, representing a 400 Gbps transceiver.

The low-energy SOA switch proposed in [35] which has an energy per bit of 15.8 pJ/b at 10 Gbps per single wavelength data rate is adopted to implement the integrated optical switch at each compute node. Since, 100 Gbps single wavelength data rate is adopted, the energy per bit of the SOA switch reduces by a factor of 10 at this data rate. As in AOPD-DCN, low power and configurable WSSs are selected as optical switches in EO-NetCoD. Each WSS has a typical operating power consumption of 50 W [36]. It is important to note that similar network setup and input parameters given in this sub-section are adopted in other sections of this paper.

TABLE I
NETWORK INPUT PARAMETERS

| Description | Value |
|---|---|
| On-board network energy per bit | 0.1 pJ/b [32] |
| Intra-rack network energy per bit | 1 pJ/b [32] |
| Electrical switch energy per bit | 5 pJ/b |
| Inter-DC gateway switch energy per bit | 10 pJ/b [32] |
| Energy per bit of routing function. | 1 pJ/b |
| Typical operating (idle) power of electrical switch | 312 W [33] |
| Typical operating power of WSS-based optical switch | 50 W [36] |
| SOA-based switch energy per bit at 100 Gbps single wavelength transmission data rate | 1.58 pJ/b |
| Single wavelength transmission data rate | 100 Gbps |

### E. *Performance Evaluation*

#### 1) *Maximum Throughput*

Under this scenario, Objective 1 of Equation (14) is adopted to maximize the throughput of all network topologies being evaluated. The results in Table II show that the maximum throughput obtained under bi-directional and unidirectional communication is equal under both variant of NetCoD. All nodes are transmitting at maximum capacity of 800 Gbps. This also confirms that interface constraints are the primary factor that determines the maximum throughput achievable. Hence, to increase total throughput, the number of wavelengths supported by compute nodes' interfaces in both variants of NetCoD must be increased. Adoption of unidirectional communication paths may also be explored to increase network throughput. However, relative to the adoption of unidirectional communication within each rack, results obtained when bi-directional communication is adopted within each rack in the DC is comparable. Hence, given appropriate intelligence in both variants of NetCoD, the size of the intra-rack backplane can be effectively halved via the adoption of bi-directional communication over a single optical link.

AOPD-DCN is also limited by interface constraints since direct point-to-point communication between compute nodes is limited to 400 Gbps. However, compute nodes also utilize the capacity of the inter-rack network to increase the overall throughput. Consequently, the maximum throughput of AOPD-DCN is greater than the throughput of both variants of NetCoD. The maximum transmitting and receiving capacity of the inter-cluster switch of AOPD-DCN is limited because a single fiber connects the ToC switch to the inter-cluster switch in the DC as shown in Fig. 1. Hence, the maximum transmitting and receiving data rate of the inter-cluster switch is capped at 800 Gbps because only limited unique (8) wavelengths are supported in the network. This limitation is easily remedied by deploying parallel optical links between the inter-cluster switch and the ToC switch in the AOPD-DCN. The adoption of a leaf-spine topology in the inter-rack of EO-NetCoD mitigates such limitation since path diversity is an inherent feature of the leaf-spine physical topology. Hence, the maximum transmitting and receiving capacity of DC gateway switch is doubled relative to that of the inter-cluster switch of AOPD-DCN.

TABLE II
MAXIMUM THROUGHPUT OF NETWORK TOPOLOGIES

| Network topology | Unidirectional | Bi-directional |
|---|---|---|
| AODP-DCN | 24.8 Tbps | N/A |
| E-NetCoD | 11.2 Tbps | 11.2 Tbps |
| EO-NetCoD | 11.2 Tbps | 11.2 Tbps |

#### 2) *Energy Efficient Network Load Test*

We further evaluate and compare the performance of both variants of NetCoD with that of AOPD-DCN by performing energy efficient network load test. The network load test considers the routing and forwarding of input traffic ($\tau_{sd}$) between nodes in the composable DC. A non-uniform traffic distribution is considered between routing and forwarding nodes in the DC. Load between 80 Gbps and 720 Gbps are considered to represent 10% to 90% utilization of each compute node's maximum throughput when a variant of NetCoD is deployed. The non-uniform traffic distribution (TD) is given by Equation (42); where Nodes A, B, C and D are compute nodes in the same rack; Node R is a randomly selected compute node in a rack remote relative to Nodes A-D; and Node G is the gateway switch of the DC.

We consider a scenario where intra-rack, inter-rack and north-south traffics accounts for 80%, 15% and 5% of each node's traffic in both directions respectively as given in Equation (42). Such traffic distribution pattern is adopted because it is expected that intra-rack traffic will be the dominant traffic within each rack in the rack-scale composable DC being considered, followed by inter-rack traffic. It is expected that north-south traffic will have the lowest percentage based on the de facto knowledge that the majority of DCN traffic is within the DC while north-south traffic accounts for a small percentage of the total traffic in the DCN.

$$TD = \begin{bmatrix} \boldsymbol{Node} & \boldsymbol{A} & \boldsymbol{B} & \boldsymbol{C} & \boldsymbol{D} & \boldsymbol{R} & \boldsymbol{G} \\ \boldsymbol{A} & 0 & 0.7 & 0.1 & 0 & 0.15 & 0.05 \\ \boldsymbol{B} & 0 & 0 & 0.7 & 0.1 & 0.15 & 0.05 \\ \boldsymbol{C} & 0.1 & 0 & 0 & 0.7 & 0.15 & 0.05 \\ \boldsymbol{D} & 0.7 & 0.1 & 0 & 0 & 0.15 & 0.05 \\ \boldsymbol{R} & 0.15 & 0.15 & 0.15 & 0.15 & 0 & 0.05 \\ \boldsymbol{G} & 0.05 & 0.05 & 0.05 & 0.05 & 0.05 & 0 \end{bmatrix} \quad (42)$$

Input traffic $\tau_{sd}$ to the MILP model is derived by multiplying the load by the traffic distribution as given in Equation (43).

$$\tau_{sd} = Load \;\; TD \qquad (43)$$

Given the input traffic $\tau_{sd}$ the energy efficiency load test is performed by adopting Objective 2 of Equation (15) to minimize the $TNPC$ as input traffic is routed and forwarded over all network topologies being evaluated. The results in Fig. 5(a) show that comparable $TNRP$ is incurred under all network topologies considered. However, marginally higher $TNRP$ is consumed under AOPD-DCN because additional routing intelligence is required to avoid wavelength collision on the link that connects the ToC switch to the gateway switch. Introducing path diversity in the inter-rack network can mitigate such limitations. The inherent path diversity of the leaf-spine physical architecture adopted in the inter-rack network of EO-NetCoD implies that such limitations are mitigated. However, the problem remains a concern. Hence, in a large deployment scenario, adoption of multiple links between switches of the inter-rack network is recommended to further leverage on path diversity via spatial multiplexing. It is important to note that the introduction of additional links between switches to achieve diversity in a practical deployment could also enhance load balancing and improve capacity and resilience. Generally, the $TNFP$ consumed by EO-NetCoD is comparable to that of AOPD-DCN as illustrated in Fig. 5(b). While the $TNFP$ increases drastically when the E-NetCoD is considered because of the many electrical switches used in the network topology.

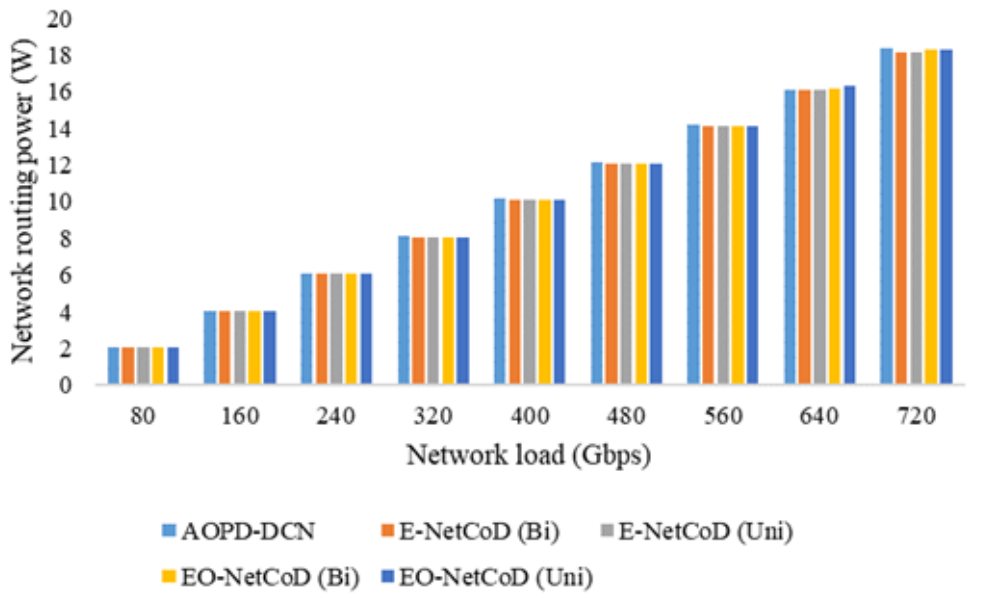

(a) Network routing power

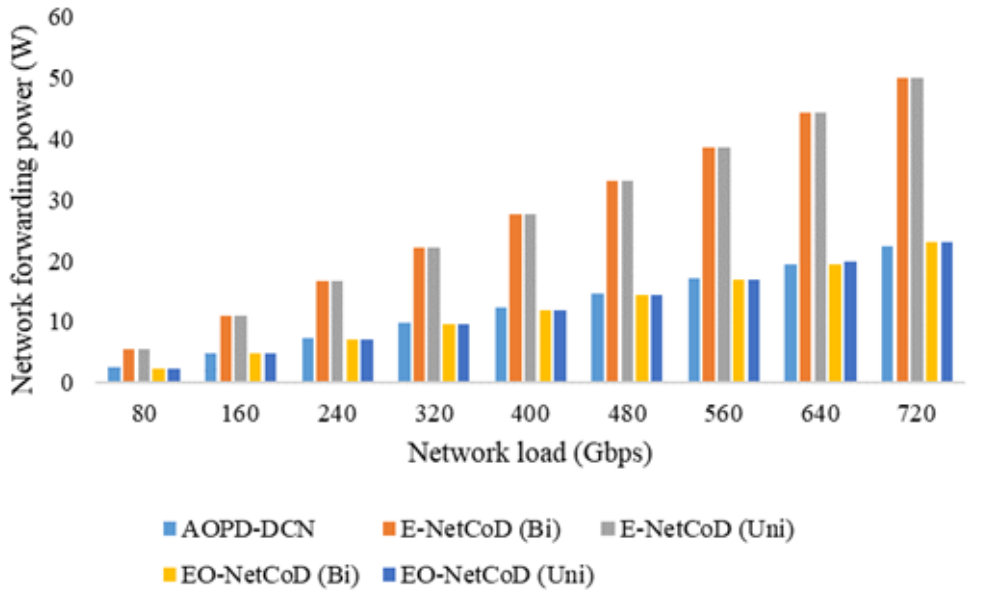

(b) Network forwarding power

Fig. 5. Routing and forwarding power of network topologies.

As expected, the power consumed by the SOA switch grows proportionally with the network loads under both variants of NetCoD as illustrated in Fig. 6(a). This is because load proportional power profile is adopted for the integrated optical switches. AOPD-DCN does not employ SOA switches, hence, SOA switch power is zero for that network topology. Furthermore, AOPD-DCN has the lowest total switch operational power (TSWOP) as shown in Fig. 6(b), because it uses one less optical switch than EO-NetCoD. On the other hand, Fig. 6(b) also shows that E-NetCoD has the higher TSWOP because it requires 6 active electrical switches. Each active electrical switch has a corresponding operating power consumption.

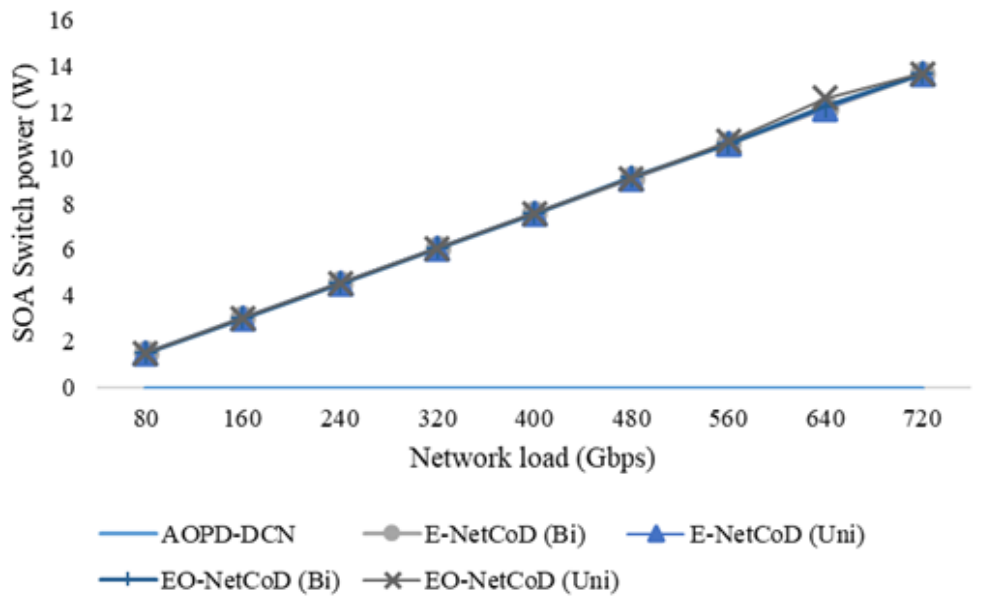

(a) SOA switch load proportional power

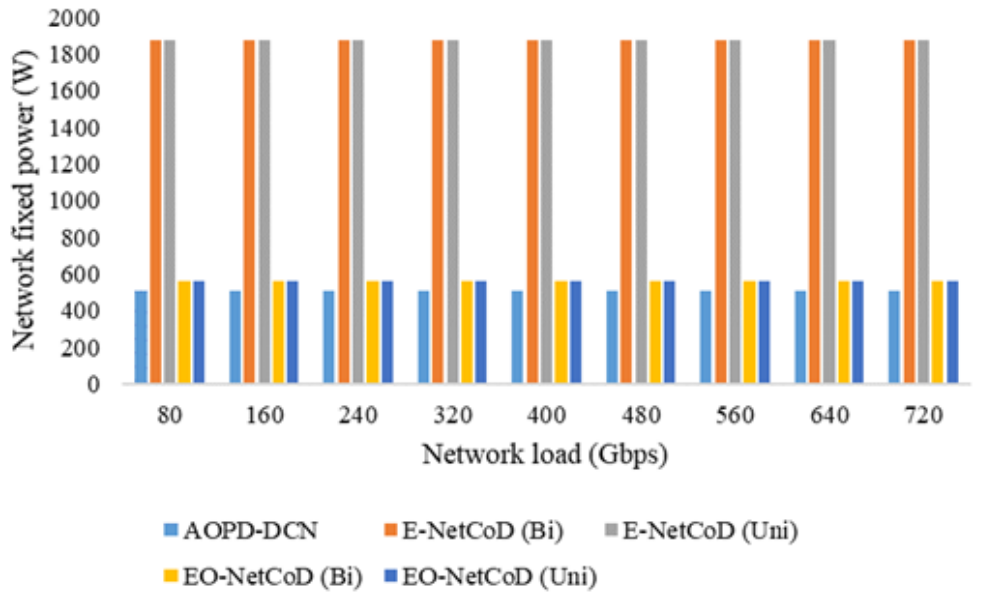

(b) Total switch operating power

Fig. 6. Switch load proportional and operating power of network topologies.

## V. MILP for Energy Efficient Placement of VMs

A MILP model that minimizes total compute power consumption, TNPC, and VM rejection in a rack-scale composable DC is given in this section. This MILP model extends the model given in Section IV. It adds compute related sets, parameters, variables, and constraints to those of the network. Like Section IV, AOPD-DCN, E-NetCoD and EO-NetCoD network topologies are adopted in the rack-scale composable DC. Furthermore, the MILP model also evaluates the performance of various forms of disaggregation in a rack-scale composable DC over different networks. Given a specific network topology and the resource demand template of each VM, the model selects the optimum placement for compute resources requested by each VM to ensure the minimization of the total compute power consumption, TNPC and the number of rejected VMs. The MILP model constraints include resource capacity constraints and resource locality constraints in addition to network constraints from Section IV. The compute related sets, parameters and variables of the MILP model are given as follows.

**Compute Related Sets and Parameters**

| | |
|---|---|
| $CR$ | Set of CPU resource components |
| $MR$ | Set of memory resource components |
| $SR$ | Set of storage resource components |
| $R$ | Set of DC racks |
| $C_j$ | Capacity of CPU component $j \in CR$ |
| IC | Idle power as a fraction of maximum CPU power consumption |
| $CPmax_j$ | Maximum power consumption of CPU component $j \in CR$ |
| $\Delta C_j$ | Power factor of CPU component $j \in CR$; $\Delta C_j = \frac{CPmax_j - IC\ CPmax_j}{C_j}$; in W/GHz |
| $M_j$ | Capacity of memory component $j \in MR$ |
| IM | Idle power consumption as a fraction of maximum memory power consumption |
| $MPmax_j$ | Maximum power consumption of memory component $j \in MR$ |
| $\Delta M_j$ | Power factor of memory component $j \in MR$; $\Delta \mathrm{M_j} = \frac{MPmax_j - IM\ MPmax_j}{M_j}$; in W/GB |
| $S_j$ | Capacity of storage component $j \in SR$ |
| IS | Idle power consumption as a fraction of maximum storage power consumption |
| $SPmax_j$ | Maximum power consumption of storage component $j \in SR$ |
| $\Delta S_j$ | Power factor of storage component $j \in SR$; $\Delta \mathrm{S_j} = \frac{SPmax_j - IS\ SPmax_j}{S_j}$; in W/GB |
| $CinN_{jn}$ | $CinN_{jn} = 1$ if CPU $j \in CR$ is placed in node $n \in N$. Otherwise $CinN_{jn} = 0$. Note that CPU components can only be placed in compute nodes. |
| $MinN_{jn}$ | $MinN_{jn} = 1$ if RAM $j \in MR$ is placed in node $n \in N$. Otherwise $MinN_{jn} = 0$. Note that memory components can only be placed in compute nodes |
| $SinN_{jn}$ | $SinN_{jn} = 1$ if hard disk drive (HDD) $j \in SR$ is placed in node $n \in N$. Otherwise $SinN_{jn} = 0$. Note that storage components can only be placed in compute nodes |
| $NinR_{nr}$ | $NinR_{nr} = 1$, If node $n \in N$ is placed in rack $r \in R$, otherwise $NinR_{nr} = 0$ |

**VM related sets and parameters**

| | |
|---|---|
| $VM$ | Set of virtual machines |
| $VCD_v$ | CPU demand of VM $v \in VM$ |
| $VMD_v$ | RAM demand of VM $v \in VM$ |
| $VSD_v$ | Storage demand of VM $v \in VM$ |
| $VCMUT_v$ | CPU to memory (RAM) traffic required by VM $v \in VM$ |
| $VCMDT_v$ | Memory (RAM) to CPU traffic required by VM $v \in VM$ |
| $VCSUT_v$ | CPU to storage traffic required by VM $v \in VM$ |
| $VCSDT_v$ | Storage to CPU traffic required by VM $v \in VM$ |
| $VMUT_v$ | Uplink north-south traffic of VM $v \in VM$ |
| $VMDT_v$ | Downlink north-south traffic of VM $v \in VM$ |
| $IMC_{sd}$ | In-memory computing traffic from VM $s \in VM$ to VM $d \in VM$ |
| $VG_{vn}$ | $VG_{vn} = 1$ denotes that node $n \in DG$ is the gateway node for north-south traffic of VM $v \in VM$ |
| $\beta$ | Cost associated with a VM rejection. |

**Variables:**

| | |
|---|---|
| $VCL_{vj}$ | $VCL_{vj} = 1$ indicates that CPU demand of VM $v \in VM$ is served by CPU $j \in CR$. Otherwise, $VCL_{vj} = 0$ |
| $VML_{vj}$ | $VML_{vj} = 1$ indicates that RAM demand of VM $v \in VM$ is served by memory (RAM) $j \in MR$. Otherwise, $VML_{vj} = 0$ |
| $VSL_{vj}$ | $VSL_{vj} = 1$ indicates that storage resource demand of VM $v \in VM$ is served by HDD $j \in SR$. Otherwise, $VSL_{vj} = 0$ |
| $CA_j$ | $CA_j = 1$ if CPU $j \in CR$ is active. Otherwise, $CA_j = 0$ |
| $MA_j$ | $MA_j = 1$ if RAM $j \in MR$ is active. Otherwise, $MA_j = 0$ |
| $SA_j$ | $SA_j = 1$ if HDD $j \in SR$ is active. Otherwise, $SA_j = 0$ |
| $CNS_n$ | $CNS_n = 1$ if compute node $n \in CN$ is active. Otherwise, $CNS_n = 0$ |
| $RS_r$ | $RS_r = 1$ if rack $r \in R$ is active. Otherwise, $RS_r = 0$ |
| $NAR$ | Number of active racks in the composable DC |
| $CM_{vsd}$ | $CM_{vsd} = 1$ if CPU resource demand of VM $v \in VM$ is placed in compute node $s \in CN$ and mermory resource demand of VM $v \in VM$ is placed in compute node $d \in CN$. Otherwise, $CM_{vsd} = 0$. |
| $CS_{vsd}$ | $CS_{vsd} = 1$ if CPU resource demand of VM $v \in VM$ is placed in compute node $s \in CN$ and storage resource demand of VM $v \in VM$ is placed in compute node $d \in CN$. Otherwise, $CS_{vsd} = 0$. |
| $MM_{sd}^{xy}$ | $MM_{sd}^{xy} = 1$ if memory to memory (in-memory computing) traffic exists from VM $x \in VM$ in compute node $s \in CN$ to VM $y \in VM$ in compute node $d \in CN$. Otherwise, $MM_{sd}^{xy} = 0$. |
| $IR_{sd}$ | Total inter-resource traffic from compute node $s \in CN$ to compute node $d \in CN$ due to VM resource demand placement. |
| $EW_{sd}$ | Total east-west traffic from node $s \in N$ to node $d \in N$. |
| $NS_{sd}$ | Total north-south traffic from node $s \in N$ to node $d \in N$. |
| $T_{sd}$ | Total traffic from node $s \in N$ to node $d \in N$. |
| $Rejected_v$ | $Rejected_v = 1$ if VM $v \in VM$ is rejected. Otherwise, $Rejected_v = 0$. |
| $TRejected$ | Total number of rejected VMs |

Certain variables in the MILP model are derived from other variables. These linear derivations form part of the linear constraints required in the MILP model. Such variables are derived as follows:

$$Rejected_v = 1 - \sum_{c \in CR} VCL_{vc} \tag{44}$$

Equation (44) derives the state of a VM using knowledge of the placement of the workload's CPU demand in any CPU component in the composable DC.

$$\sum_{v \in VM} VCL_{vc} \geq CA_c \quad \forall c \in CR \tag{45}$$

$$\sum_{v \in VM} VCL_{vc} \leq Q\, CA_c \quad \forall c \in CR \tag{46}$$

$$\sum_{v \in VM} VML_{vm} \geq MA_m \quad \forall m \in MR \tag{47}$$

$$\sum_{v \in VM} VML_{vm} \leq Q\, MA_m \quad \forall m \in MR \tag{48}$$

$$\sum_{v \in VM} VSL_{vs} \geq SA_s \quad \forall s \in SR \tag{49}$$

$$\sum_{v \in VM} VSL_{vs} \leq Q\, SA_s \quad \forall s \in SR \tag{50}$$

Equations (45) – (50) derive the state of CPU, memory, and storage resources components. The state of each resource component depends on the utilization of each resource type to satisfy resource demands of any active VM.

**Total CPU Power Consumption**

Total CPU power consumption ($TCPC$) in the DC is derived as follows.

$$TCPC = \sum_{c \in CR} \left( IC\, CPmax_c\, CA_c + \sum_{v \in VM} \Delta C_c\, VCL_{vc}\, VCD_v \right) \tag{51}$$

**Total Memory Power Consumption**

Total memory power consumption ($TMPC$) in the DC is derived as follows.

$$TMPC = \sum_{m \in MR} \left( IM\, MPmax_m\, MA_m + \sum_{v \in VM} \Delta M_m\, VML_{vm}\, VMD_v \right) \tag{52}$$

**Total Storage Power Consumption**

Total storage power consumption ($TSPC$) in the DC is derived as follows.

$$TSPC = \sum_{s \in SR} \left( IS\, SPmax_s\, SA_s + \sum_{v \in VM} \Delta S_s\, VSL_{vs}\, VSD_v \right) \tag{53}$$

**Total Compute Power Consumption**

Total compute power consumption ($TComPC$) in the DC is derived as follows.

$$TComPC = TCPC + TMPC + TSPC \tag{54}$$

**Derived Network Variables**

$$\sum_{v \in VM} \sum_{c \in CR} VCL_{vc}\, CinN_{cn} + \sum_{v \in VM} \sum_{m \in MR} VML_{vm}\, MinN_{mn} + \sum_{v \in VM} \sum_{s \in SR} VSL_{vs}\, SinN_{sn} \geq CNS_n \quad \forall n \in CN \tag{55}$$

$$\sum_{v \in VM} \sum_{c \in CR} VCL_{vc}\, CinN_{cn} + \sum_{v \in VM} \sum_{m \in MR} VML_{vm}\, MinN_{mn} + \sum_{v \in VM} \sum_{s \in SR} VSL_{vs}\, SinN_{sn} \leq Q\, CNS_n \quad \forall n \in CN \tag{56}$$

Equations (55) and (56) derive the state of each compute node based on the use of CPU, memory, or storage resource in that compute node to satisfy the resource demand of any VM.

$$\sum_{n \in CN} CNS_n\, NinR_{nr} \geq RS_r \quad \forall r \in R \tag{57}$$

$$\sum_{n \in CN} CNS_n\, NinR_{nr} \leq Q\, RS_r \quad \forall r \in R \tag{58}$$

Equations (57) and (58) derive the state of each rack based on the state of compute nodes in that rack.

$$NAR = \sum_{r \in R} RS_r \tag{59}$$

Equation (59) derives the number of active racks in the DC.

$$CM_{vsd} \leq \sum_{c \in CR} VCL_{vc}\, CinN_{cs} \quad \forall v \in VM, s \in N, d \in N \tag{60}$$

$$CM_{vsd} \leq \sum_{m \in MR} VML_{vm}\, MinN_{md} \quad \forall v \in VM, s \in N, d \in N \tag{61}$$

$$CM_{vsd} \geq \sum_{c \in CR} VCL_{vc}\, CinN_{cs} + \sum_{m \in MR} VML_{vm}\, MinN_{md} - 1 \quad \forall v \in VM, s \in N, d \in N \tag{62}$$

Equations (60) - (62) implement the product of two derived binary variables as illustrated in Equation (63).

$$CM_{vsd} = \sum_{c \in CR} VCL_{vc}\, CinN_{cs} \sum_{m \in MR} VML_{vm}\, MinN_{md} \quad \forall v \in VMv \in VM, s \in N, d \in N \tag{63}$$

Equation (63) derives $CM_{vsd}$ which gives the compute nodes where the CPU and memory demands of VM $v \in VM$ are placed.

$$CS_{vsd} \leq \sum_{c \in CR} VCL_{vc}\, CinN_{cs} \quad (64)$$
$$\forall v \in VM, s \in N, d \in N$$

$$CS_{vsd} \leq \sum_{n \in SR} VSL_{vn}\, SinN_{nd} \quad (65)$$
$$\forall v \in VM, s \in N, d \in N$$

$$CS_{vsd} \geq \sum_{c \in CR} VCL_{vc}\, CinN_{cs} + \sum_{n \in SR} VSL_{vn}\, SinN_{nd} - 1 \quad (66)$$
$$\forall v \in VM, s \in N, d \in N$$

Equations (64) - (66) implement the product of two derived binary variables as illustrated in Equation (67).

$$CS_{vsd} = \sum_{c \in CR} VCL_{vc}\, CinN_{cs} \sum_{n \in SR} VSL_{vn}\, SinN_{nd} \quad (67)$$
$$\forall v \in VM, s \in N, d \in N$$

Equation (67) derives $CS_{vsd}$ which specifies the compute nodes where the CPU and storage demands of VM $v \in VM$ are placed.

$$MM_{sd}^{xy} \leq \sum_{m \in MR} VML_{xm}\, MinN_{ms} \quad (68)$$
$$\forall x, y \in VM, s \in N, d \in N$$

$$MM_{sd}^{xy} \leq \sum_{m \in MR} VML_{ym}\, MinN_{md} \quad (69)$$
$$\forall x, y \in VM, s \in N, d \in N$$

$$MM_{sd}^{xy} \geq \sum_{m \in MR} VML_{xm}\, MinN_{ms} + \sum_{m \in MR} VML_{ym}\, MinN_{md} - 1 \quad (70)$$
$$\forall x, y \in VM, s \in N, d \in N$$

Equations (68) - (70) implement the product of two derived binary variables as illustrated in Equation (71).

$$MM_{sd}^{xy} = \sum_{m \in MR} VML_{xm}\, MinN_{ms} \sum_{m \in MR} VML_{ym}\, MinN_{md} \quad (71)$$
$$\forall x, y \in VM, s \in N, d \in N$$

Equation (71) derives $MM_{sd}^{xy}$ which gives a VM pair $(x, y)$ that exchanges in-memory computing traffic via memory-to-memory communication and the corresponding compute nodes ($s$ and $d$) where the memory components, which support the memory demand of each VM in the pair, are placed.

$$IR_{sd} = \sum_{v \in VM} (CM_{vsd}\, VCMUT_v + CM_{vds}\, VCMDT_v + CS_{vsd}\, VCSUT_v + CS_{vds}\, VCSDT_v) \quad (72)$$
$$\forall s, d \in CN$$

Equation (72) derives $IR_{sd}$ which is the inter-resource traffic between node $s \in N$ and node $d \in N$. Resource locality constraints in Equations (81) and (82) ensure that nodes $s$ and $d$ are always in the same rack. Furthermore, resource allocation ensures that source and destination nodes ($s$ and $d$) are compute nodes.

$$EW_{sd} = \sum_{x \in VM} \sum_{y \in VM: x \neq y} MM_{sd}^{xy}\, IMC_{xy} \quad (73)$$
$$\forall s, d \in CN$$

Equation (73) derives $EW_{sd}$ which is the east-west traffic between memory components as a result of the placement of memory demands of VMs.

$$NS_{sd} = \sum_{v \in VM} \sum_{c \in CR} (VCL_{vc}\, CinN_{cs}\, VMUT_v\, VG_{vd} + VCL_{vc}\, CinN_{cd}\, VMDT_v\, VG_{vs}) \quad (74)$$
$$\forall s, d \in CG$$

Equation (74) derives $NS_{sd}$ which is the north-south traffic from node $s \in CG$ to node $d \in CG$ in the DC. Note that the source of south-bound traffic is a gateway switch in the DC. A gateway switch is also the destination of north-bound traffic in the DC.

$$T_{sd} = IR_{sd} + EW_{sd} + NS_{sd} \quad (75)$$
$$\forall s, d \in CG$$

Equation (75) derives the traffic demand to be routed and forwarded over the composable DCN.

$$TOBP = \sum_{s \in CN} \sum_{d \in CN: s = d} 2IR_{sd}\, OBepb + \sum_{s \in CN} \sum_{d \in CG: s \neq d} T_{sd}\, OBepb + \sum_{d \in CN} \sum_{s \in CG: s \neq d} T_{sd}\, OBepb \quad (76)$$

Equation (76) derives the total on-board power ($TOBP$) consumption due to the traversal of the on-board fabric by internal, ingress and egress traffic of all compute nodes.

The $TNPC$ from Equation (13) is revised to include the TOBP as follows in Equation (77).

$$TNPC = TNFP + TOBP + TNRP + TXNP \quad (77)$$

The model for energy efficient placement of VM in rack-scale composable DC is defined as follows:

**Objective 3**: Minimize:

$$TComPC + TNPC + \beta\, TRejected \quad (78)$$

Equation (78) is the objective of the model for energy efficient placement of VMs in rack-scale composable DC. It minimizes the total compute and network power consumption and the number of rejected VMs. β is the cost (measured in Watts) associated with each rejected VM. $\beta \gg 1$ denotes that high cost is associated with each rejected VM.

**Subject to**:

**Compute constraints**

$$\sum_{j \in CR} VCL_{vj} \leq 1 \quad (79)$$
$$\forall v \in VM$$

$$\sum_{j \in CR} VCL_{vj} = \sum_{j \in MR} VML_{vj} \quad (80)$$
$$\forall v \in VM$$

$$\sum_{j \in CR} VCL_{vj} = \sum_{j \in SR} VSL_{vj} \quad (81)$$
$$\forall v \in VM$$

Constraints (79) - (81) limit the maximum number of nodes that can host CPU, memory, and storage resource demands of a VM to one. This is because neither replication nor slicing of workloads is permitted. The constraints also permit VM rejection in scenarios where resource capacity is limited. The constraints ensure that the VM is fully embedded.

$$\sum_{n \in N} \sum_{c \in CR} VCL_{vc}\, CinN_{cn}\, NinR_{nr} = \sum_{n \in N} \sum_{m \in MR} VML_{vm}\, MinN_{mn}\, NinR_{nr} \quad \forall v \in VM, r \in R \tag{82}$$

$$\sum_{n \in N} \sum_{c \in CR} VCL_{vc}\, CinN_{cn}\, NinR_{nr} = \sum_{n \in N} \sum_{s \in SR} VSL_{vs}\, SinN_{sn}\, NinR_{nr} \quad \forall v \in VM, r \in R \tag{83}$$

Constraints (82) and (83) are the locality constraints of rack-scale composable DC. They ensure that CPU, memory, and storage components used to provision a given VM are in the same rack but not necessarily in the same compute node.

$$\sum_{v \in VM} VCD_v\, VCL_{vc} \le C_c \quad \forall c \in CR \tag{84}$$

$$\sum_{v \in VM} VMD_v\, VML_{vm} \le M_m \quad \forall m \in MR \tag{85}$$

$$\sum_{v \in VM} VSD_v\, VSL_{vs} \le S_s \quad \forall s \in SR \tag{86}$$

Constraints (84) - (86) denote resource capacity constraints for each CPU, memory, and storage components. These equations also ensure that a compute resource's capacity is conserved.

## VI. Energy Efficient Placement of VMs

Using the combined MILP models from Section IV and Section V, energy efficient placement of VMs in rack-scale composable DCs is studied. Rack-scale DCs that implement logical, hybrid and physical disaggregation are considered when a zero-cost-and-un-capacitated network and non-zero-cost-and-capacitated networks (i.e., AOPD-DCN, E-NetCoD or EO-NetCoD) are deployed. Two classes of CPU, memory, and storage resource components (illustrated in Table III) are considered to reflect the heterogeneity of compute resources deployed in production DCs. To study the performance of all network topologies within a rack, a composable DC with a single rack is considered in this section. Allocated to the single rack are 8 CPU, 8 memory and 8 storage resource components which are distributed into compute nodes according to the disaggregation approach adopted as shown in Fig. 7.

There are eight heterogenous compute nodes in the single rack of a logically disaggregated composable DC and each heterogeneous compute node comprises of one CPU, one memory and one storage component as shown in Fig. 7(a). The single rack comprises of 10 compute nodes when hybrid disaggregation is adopted in the DC as shown in Fig. 7(b). Compute nodes 1 – 4 are heterogeneous nodes, each comprising of one CPU, one memory and one storage component. Compute nodes 5 – 10 are homogenous, each homogenous node comprises of two CPU or two memory or two storage components of the same component class. The single rack in a physically disaggregated DC comprises of 12 homogenous compute nodes and each compute node comprise of two CPU or two memory or two storage components from the same class as shown in Fig. 7(c). Because the MILP model's complexity grows as the number network nodes increases, we further simplify the MILP model in single rack scenario by excluding spine switches or ToC switches from corresponding network topologies. A scenario where the ToR switch is connected directly to the DC gateway switch is considered. Furthermore, the network parameters from Section IV.D are adopted in this section.

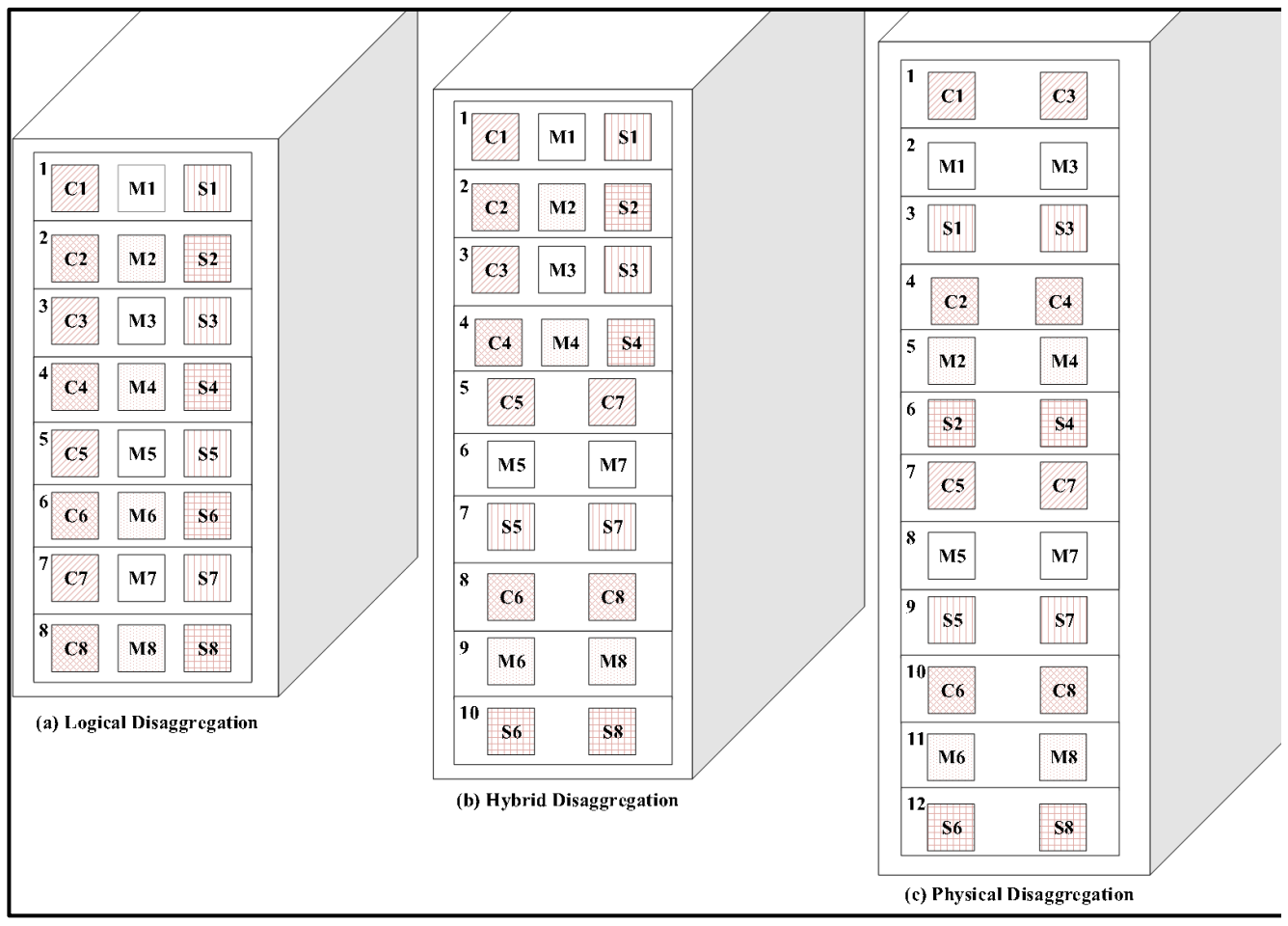

Fig. 7. Resource disaggregation of single rack.

TABLE III
Compute Component Capacity and Peak Power

| ID | CPU Capacity (Peak Power) | RAM Capacity (Peak Power) | HDD Capacity (Peak Power) |
|---|---|---|---|
| 1 | 3.6 GHz (130 W) | 32 GB (40 W) | 320 GB (6.19 W) |
| 2 | 2.66 GHz (95 W) | 24 GB (30.72 W) | 250 GB (6.19 W) |
| 3 | 3.6 GHz (130 W) | 32 GB (40 W) | 320 GB (6.19 W) |
| 4 | 2.66 GHz (95 W) | 24 GB (30.72 W) | 250 GB (6.19 W) |
| 5 | 3.6 GHz (130 W) | 32 GB (40 W) | 320 GB (6.19 W) |
| 6 | 2.66 GHz (95 W) | 24 GB (30.72 W) | 250 GB (6.19 W) |
| 7 | 3.6 GHz (130 W) | 32 GB (40 W) | 320 GB (6.19 W) |
| 8 | 2.66 GHz (95 W) | 24 GB (30.72 W) | 250 GB (6.19 W) |

Since, the complexity of the model also grows with the number of VMs and size of the composable DC, only few VMs are considered for placement in the single rack. We consider 12 input VMs with a mix of compute demand intensity as illustrated in Table IV. Furthermore, the data rate of each VM's CPU-memory communication, CPU-disk communication and north-south communication are as also given in Table IV. Traffic demands of VMs in Table IV are generated via uniform distribution of total CPU-memory traffic, CPU-storage traffic, and CPU-IO traffic over specific ranges. VMs are clustered into in-memory communication groups (IMCG), as illustrated in Table IV, to represent sets of related VMs in conventional DCs. Each set of related VMs have one-to-one, one-to-many, many-to-many, or mixed in-memory computing traffic patterns between the applications of the group. The range of in-memory computing traffic between two VMs in the same group is 5 Gbps to 40 Gbps.

The MILP model is solved using the 64-bit AMPL/CPLEX solver on the ARC3 supercomputing node in University of Leeds with 24 CPU cores and 128 GB of memory [37]. Our analysis of results from the model focuses on metrics such as total computing power consumption, total network power consumption, number of active compute components, average active compute component utilization and network available throughput utilization. To obtain optimal results, it turns out that the MILP model bin-packs VMs compute demands into compute components to achieve optimal power and utilization efficiencies within compute capacity constraints and network constraints.

TABLE IV
VM COMPUTE AND NETWORK DEMANDS

| VM ID | CPU demand (GHz) | Memory demand (GB) | Storage demand (GB) | CPU to RAM (Gbps) | RAM to CPU (Gbps) |
|---|---|---|---|---|---|
| 1 | 1.8 | 7.2 | 80 | 116.7 | 50 |
| 2 | 1.8 | 24 | 240 | 50 | 66.7 |
| 3 | 2.6 | 10.8 | 120 | 100 | 41.7 |
| 4 | 0.9 | 13 | 160 | 266.7 | 116.7 |
| 5 | 0.9 | 3.6 | 160 | 466.7 | 100 |
| 6 | 2.6 | 32 | 160 | 466.7 | 50 |
| 7 | 1.8 | 24 | 80 | 333.3 | 44.4 |
| 8 | 2.6 | 10.8 | 80 | 133.3 | 233.3 |
| 9 | 2.6 | 32 | 80 | 166.7 | 66.7 |
| 10 | 1.8 | 7.2 | 160 | 116.7 | 100 |
| 11 | 1.8 | 24 | 240 | 433.3 | 66.7 |
| 12 | 2.6 | 10.8 | 80 | 333.3 | 25 |
| VM ID | In-memory communication group | CPU to HDD (Gbps) | HDD to CPU (Gbps) | CPU to IO (Gbps) | IO to CPU (Gbps) |
| 1 | A | 26 | 9.3 | 9.5 | 6.6 |
| 2 | A | 60 | 9.5 | 3.3 | 4 |
| 3 | A | 64 | 6 | 5 | 3.5 |
| 4 | A | 86 | 5 | 3 | 4 |
| 5 | A | 23 | 9 | 10 | 2 |
| 6 | B | 20 | 28 | 2.6 | 2.5 |
| 7 | B | 64 | 19.5 | 2.75 | 3.5 |
| 8 | B | 17.5 | 14 | 1.7 | 5.7 |
| 9 | B | 10 | 29 | 1 | 8.5 |
| 10 | B | 14 | 49 | 2 | 3 |
| 11 | C | 68 | 45 | 1.75 | 3 |
| 12 | C | 22 | 9.7 | 4 | 2 |

### *A. Zero-Cost-and-Un-capacitated Network*

Under this scenario, energy efficient placement of VMs is performed in rack-scale DCs that employ logical, physical or hybrid disaggregation over an un-capacitated-and-zero-cost network. The zero-cost-and-un-capacitated network has no network capacity constraints, and zero power is consumed as a result of traffic forwarding and routing over the network. The results in Fig. 8, and Fig. 9 show that all disaggregation approaches achieved optimal efficiencies as compute components are utilized based on available capacity and energy efficiency. VM compute demands are bin-packed into CPU, memory, and storage components to avoid VM rejection and to achieve optimal energy efficiency under the corresponding form of disaggregation employed in the rack. Since any form of network cost is absent under this scenario, optimal compute energy efficiency is achieved under all forms of disaggregation employed in the single rack. Hence, the utilization of active compute components is maximized within the component's available capacity and their corresponding energy efficiency. Fig. 8(a) shows that equal number of CPU, memory and storage components are activated under all forms of disaggregation employed in the DC. While all (8) CPU and memory components in the rack are activated to prevent VM rejection, only 6 storage components are activated. This is because the storage demands of VMs as given in Table IV are less intensive relative to the capacity of storage components considered. Hence, to promote greater energy efficiency, consolidation of storage demands into 320 GB storage components is preferred while fewer 250 GB storage components are activated as shown in Fig. 8(a). This is because the 320 GB storage component is more energy efficient than the 250 GB storage component as it can support higher capacity at the same peak power consumption as the 250 GB storage component.

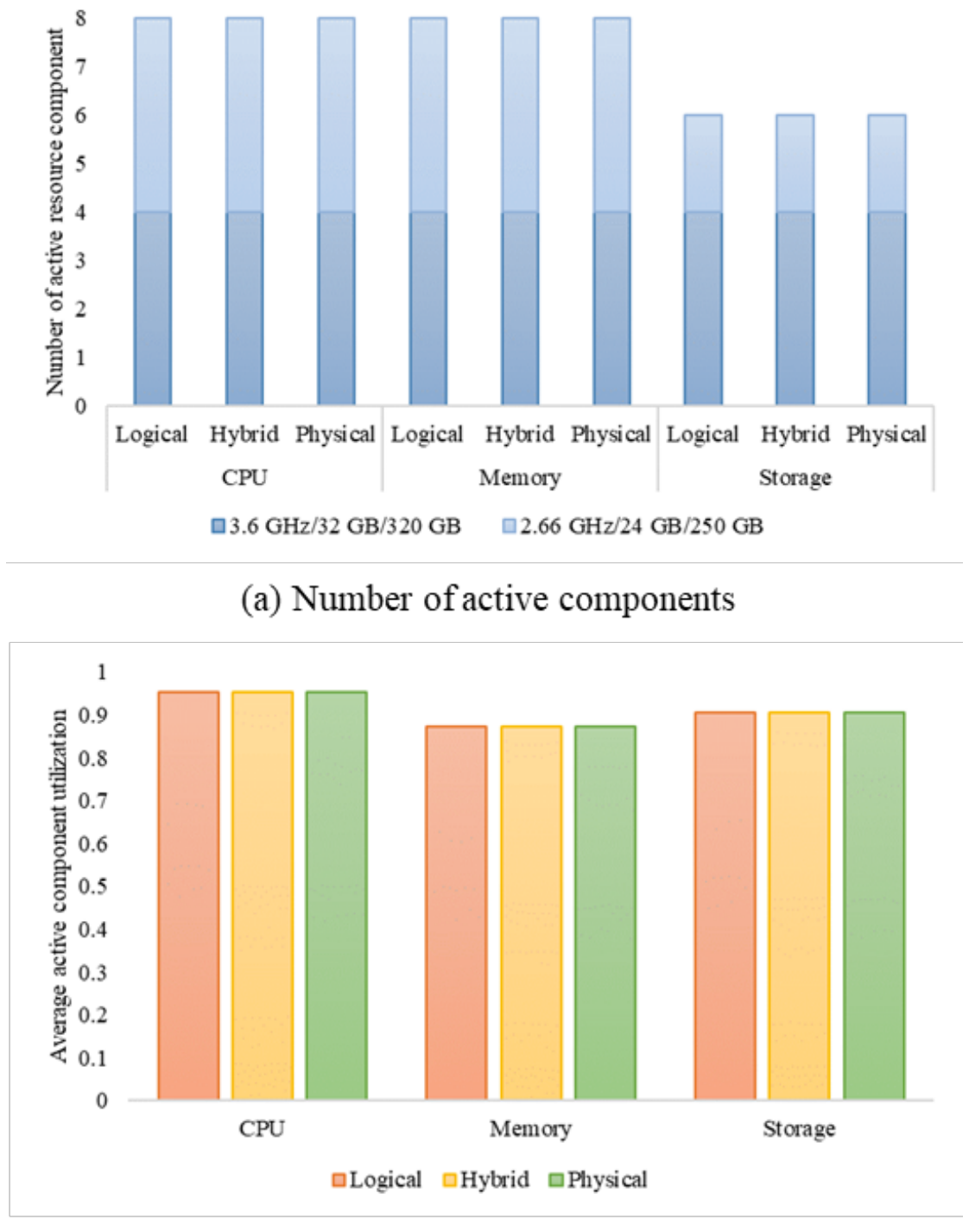

(a) Number of active components

(b) Average utilization of active components

Fig. 8. Number and average utilization of active compute components under zero-cost-and-un-capacitated network.

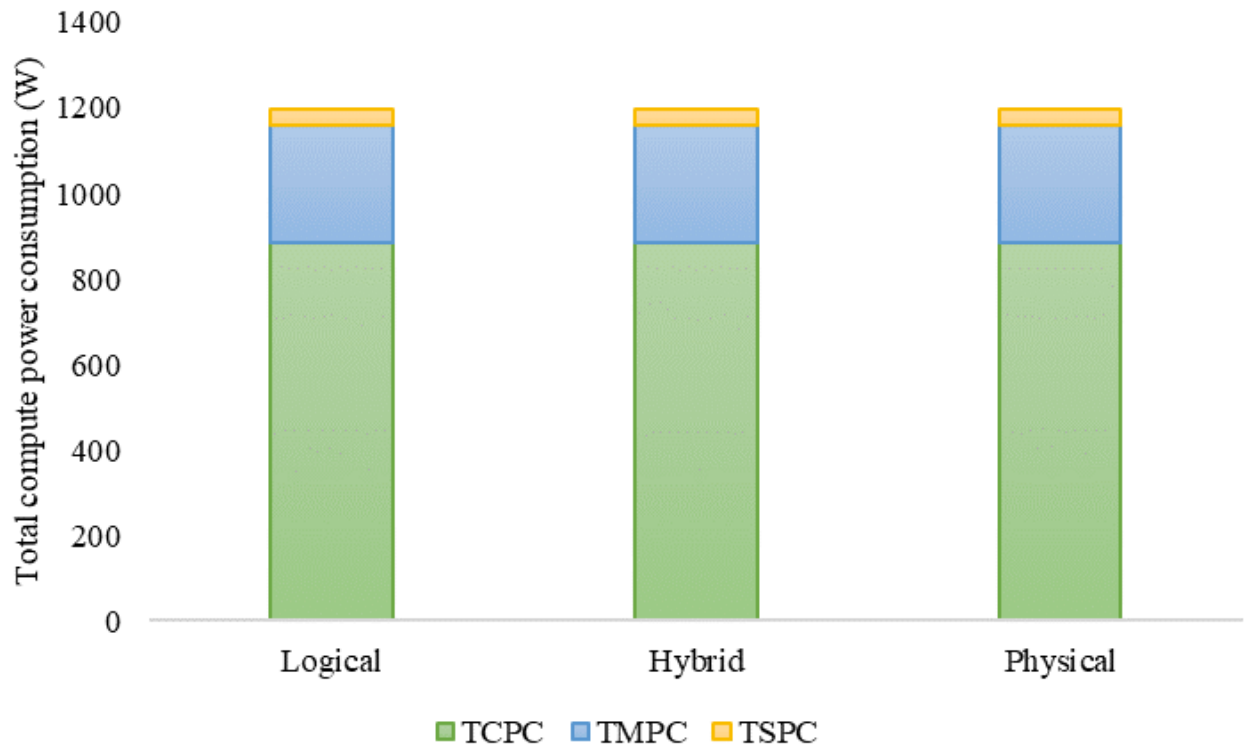

Fig. 9. Total compute power consumption under zero-cost-and-un-capacitated network.

Furthermore, equal average active compute component utilization is obtained when all forms of disaggregation are employed in the rack-scale DC as shown in Fig. 8(b). As a consequent of optimal utilization of CPU, memory, and storage components under all forms of disaggregation employed in the rack, equal TCPC, TMPC and TSPC are also obtained under all disaggregation types considered as shown in Fig. 9. An important observation is that having obtained the same results under all disaggregation approaches does not always imply that the placement of VM compute demands are also the same under different forms of disaggregation. Hence, as is the case in our scenarios, different VM placements may achieve the desired optimal performance if the compute component capacity constraint is satisfied. The results obtained under this scenario demonstrate the efficacy of all forms of disaggregation to achieve optimal efficiency when an arbitrary un-capacitated-and-zero-cost network is available. However, it is expected that the results will change when network constraints and cost are present in the composable DCs.

### *B. Non-Zero-Cost-and-Capacitated Networks*

A non-zero-cost-and-capacitated network is a network with non-zero-power consumption and network capacity constraints. As defined in the objective function of the MILP model, there are three primary factors that influence placement of VM demands in non-zero-cost networks. These are VM rejection, compute energy efficiency and TNPC. However, because a high cost is associated with the rejection of all VMs in the DC, rejection of VMs is strongly discouraged. Hence, to achieve optimal results when a non-zero-cost-and-capacitated network is employed in the rack-scale DC, best-effort will be made to prevent VM rejection while trade-offs between compute energy efficiency and network energy efficiency are also considered. Placement of VM demands is also expected to be constrained by both compute and network constraints stated in the MILP model.

#### *1) Logical Disaggregation*

When logical disaggregation is considered in the rack-scale composable DC, similar VM placement is replicated for all network topologies considered. Hence, the optimal placement of VM demands does not change with the network topology that is adopted. However, relative to the zero-cost-and-un-capacitated network, the placement of VMs when AOPD-DCN, E-NetCoD or EO-NetCoD is deployed is sub-optimal. This is because of the presence of network constraints and the introduction of network power consumption. Even though, the resulting traffic matrix generated under un-capacitated-and-zero-cost network can be routed via AOPD-DCN, the additional network power that must be consumed to achieve the same compute energy efficiency as the zero-cost network scenario outweighs the potential benefits that could be achieved. Therefore, an alternative VM placement strategy is adopted when AOPD-DCN is deployed to minimize TNPC. On the other hand, the resulting traffic generated under the zero-cost-and-un-capacitated network scenario is unrouteable by both variants of NetCoD because of the interface capacity constraint at compute nodes. Hence, an alternative VM placement strategy is required to satisfy network constraints and to minimize TNPC when both variants of NetCoD are deployed.

A strategy that balances the trade-offs between compute power consumption and TNPC is adopted to obtain optimal placement. Energy efficient placement of CPU demands is often given higher priority. Hence, CPU demands are consolidated (within resource capacity constraint as much as possible) to achieved high utilization of active CPU components. However, to ensure that TNPC is minimized, memory demands of some VMs are placed in the same compute node as the CPU demand. This strategy is strongly applied to VMs that are known to have very high total CPU-to-memory traffic. It is also applied to some VMs that are known to have moderately high CPU-to-memory traffic, provided that the capacity of the memory component in the corresponding compute node is enough. Otherwise, the memory demands of such VMs are placed in different nodes to achieve lower TMPC by ensuring high utilization of active memory components. The memory demand of VMs that have low-medium volume of total CPU-to-memory traffic are often provisioned in memory components that are in a different compute node from the CPU component that hosts the CPU demand. This strategy is adopted to ensure high utilization of active memory components. Consequently, TMPC is also minimized. The adopted strategy also reduces the impact of disaggregation on the network by minimizing the TNPC because a lower volume of traffic is sent over the network.

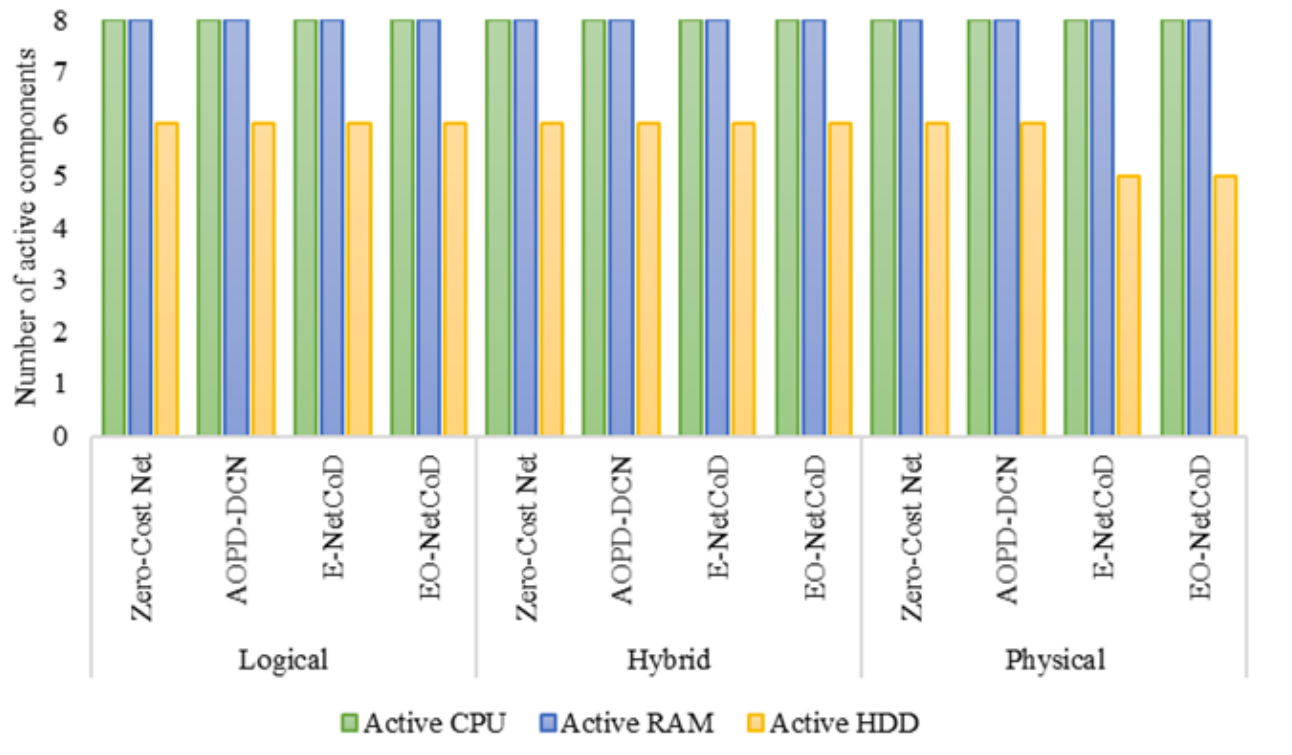

(a) Number of active components

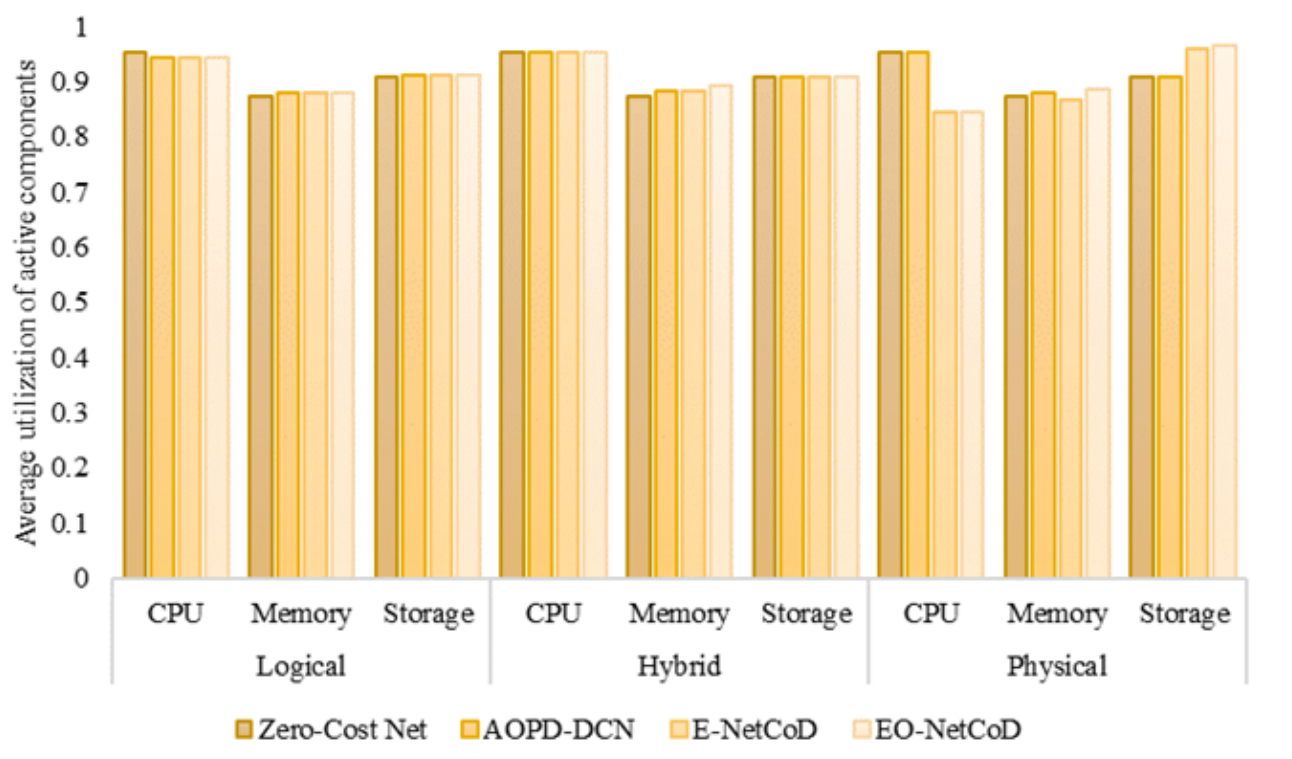

(b) Average utilization of active components

Fig. 10. Number and average utilization of active compute components.

Additionally, TNPC is further minimized by placing the storage demand of most VMs in the same compute node as the CPU demand of the VM. This reduces the volume of traffic in the network and the consequential power that would have been consumed. However, the storage demand of some VMs is also placed in a different compute node from the node hosting the VM's CPU demand to reduce the TSPC by optimally utilizing active storage components. This also reduces the number of active storage components as illustrated in Fig.10(a).

It is also important to note that, network traffic is also minimized by reducing or eliminating in-memory communication between VMs in the same IMCG. This is done by placing such VMs into the same memory component. For instance, the memory demand of VMs 3, 4, and 5 which belong to the IMCG-A are always placed in the same memory component. As a result, in-memory communication between such VMs is avoided since the CPU allocated to each VM can be granted access to the appropriate address on the common (shared) memory component to retrieve data. However, because in-memory communication volume is relatively small, the reduction of the volume of in-memory communication in the network does not have a strong effect on VM demand placement decisions.

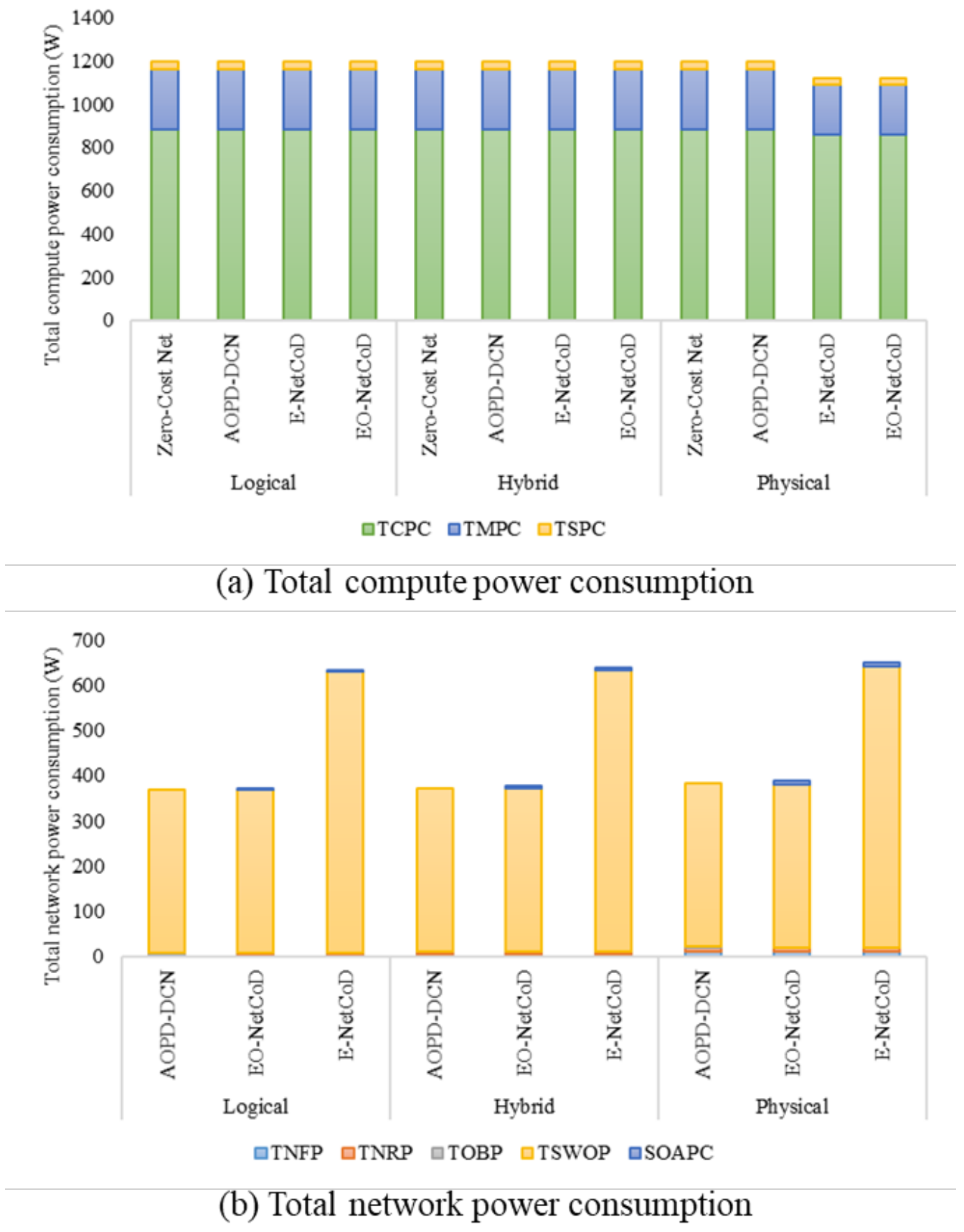

Fig. 11. Compute and network power consumption in composable DC.

Relative to the zero-cost-and-un-capacitated network, the total compute (CPU, memory, and storage) power consumption increased marginally when non-zero-cost-and-capacitated networks are deployed in a logically disaggregated rack. Marginal increase in TCPC, TMPC and TSPC are responsible for the marginal increase in total compute power consumption. The less efficient placement of VM demands when AOPD-DCN, E-NetCoD or EO-NetCoD is deployed is responsible for the fall in average utilization of active CPU and storage components as shown in Fig. 10(b). The fall in the average utilization of active CPU components is somewhat marginal. Compared to the zero-cost-and-un-capacitated network scenario, the average utilization of active memory and storage components marginally increases as shown in Fig. 10(b). However, the marginal increase in average utilization of active memory and storage components does not result in a fall in TMPC and TSPC respectively because it is achieved by increasing the utilization of memory and storage components that are less energy efficient. Hence, the TMPC and TSPC increased marginally in-spite of the increase in average active memory and storage utilization.

The results obtained under the logical disaggregation setup provide an excellent basis for further fair comparison of the various network topologies being considered. The TNRP is the same for AOPD-DCN and both variants of NetCoD. This is because the traffic demand between two communicating nodes pair is always routed directly between the two nodes in the logical layer of all network topologies considered. Hence, all end-to-end logical paths created over the physical topologies (AOPD-DCN, E-NetCoD and EO-NetCoD) are direct and do not employ intermediated routing nodes along the path. In the physical layer, the TNFP consumed by AOPD-DCN and EO-NetCoD are equal. This is because the optical ToR switch is the only intermediate node traversed by end-to-end light paths created over the physical topology. Since, the optical switch only has a constant (low) operating power consumption that is non-load proportional, it has no impact on the TNFP of both AOPD-DCN and EO-NetCoD. On the other hand, the E-NetCoD topology employs an electrical switch which has both fixed operational and load proportional components in its power profile. Hence, relative to the AOPD-DCN or EO-NetCoD topologies, the E-NetCoD topologies has higher TNFP since the ToR switch is an important intermediate node traversed by some direct logical layer links created between communicating nodes pairs.

The TOBP is the same across all three topologies because the VMs are placed in the same way under all topologies. SOAPC makes no contribution to the TNPC of AOPD-DCN since the SOA switches are not required at each compute node. On the other hand, the same SOAPC is obtained under both variants of NetCoD; hence, the contribution of SOAPC to the TNPC is the same as shown in Fig. 11(b). The TSWOP of AOPD-DCN and EO-NetCoD are equal as both topologies employed one optical switch and one electrical gateway switch with equal typical operating power consumption. The TSWOP of the E-NetCoD is higher as illustrated in Fig. 11(b) because two electrical switches are required when the topology is deployed in this evaluation scenario.

To achieve a balanced trade-off between compute power consumption and TNPC under all network topologies considered, the optimal VM placement obtained via MILP optimization enabled zero-hop communication between all intra-rack communicating node pairs. Furthermore, single hop communication is employed for communication between compute nodes and the gateway switch. Hence, the SOAPC and/or the TNRP and TNFP consumed due to VMs placement are minimized under the corresponding topology. The TNPC increases by 0.7% when AOPD-DCN is replaced by EO-NetCoD. The energy efficient SOA switches employed in EO-NetCoD are solely responsible for the marginal increase in TNPC observed. The TNPC increased by 71% when EO-NetCoD is replaced by E-NetCoD. The TSWOP and the TNFP consumed by the electrical ToR switch, which is an important intermediate node, are responsible for this relative increase in TNPC as shown in Fig. 11(b).

However, evaluation of power consumption alone is not sufficient to evaluate network performance. The AOPD-DCN topology adopts a generic design to achieve a suitable composable DCN. Hence, in a logically disaggregated DC that employs AOPD-DCN, each compute node requires multiple dedicated interfaces to ensure full mesh physical connectivity in the rack. This becomes a design problem as the number of nodes in each rack increases. This is because multiple (up to 48 high-capacity) interfaces must be fitted onto each compute node. Both variants of NetCoD mitigate this problem by adopting a targeted design that addresses the specific challenge posed by resource disaggregation in a practical composable DC.

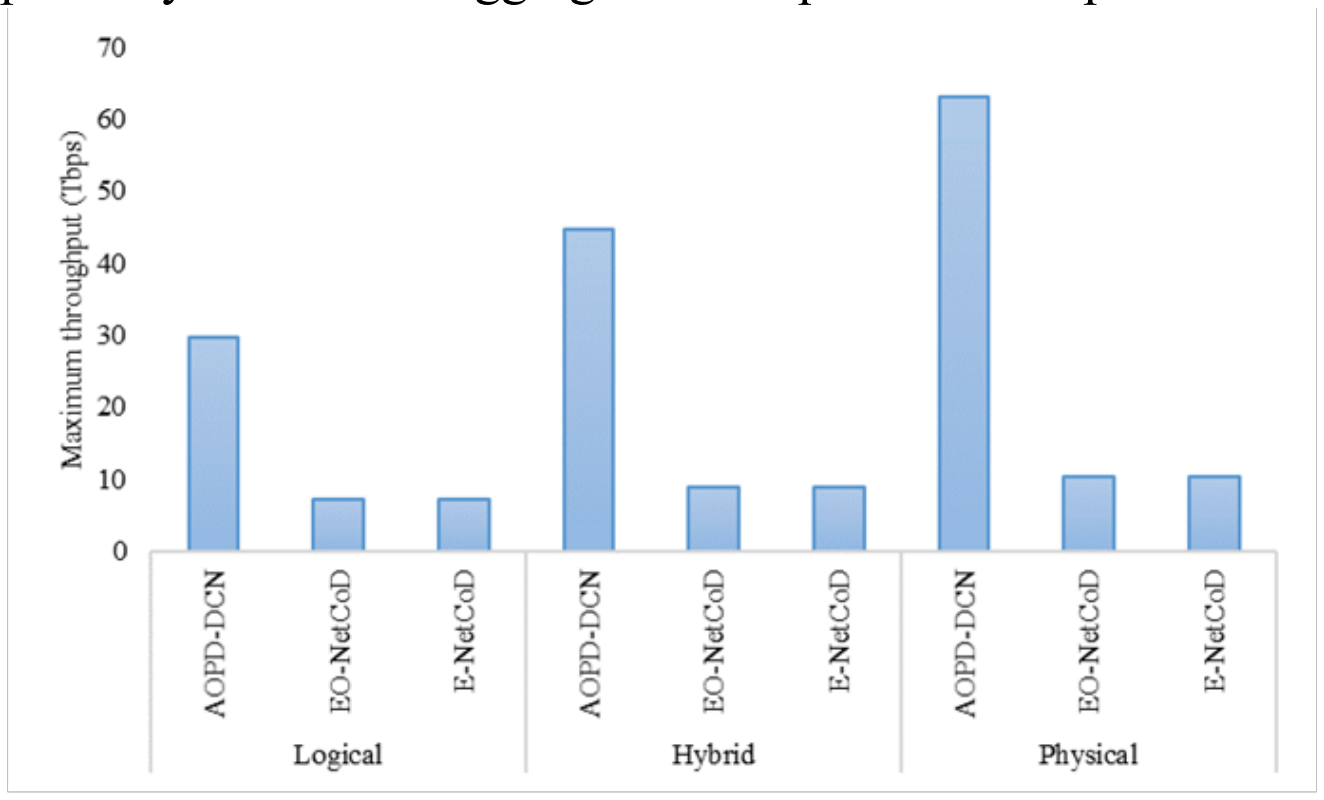

(a) Maximum network throughput

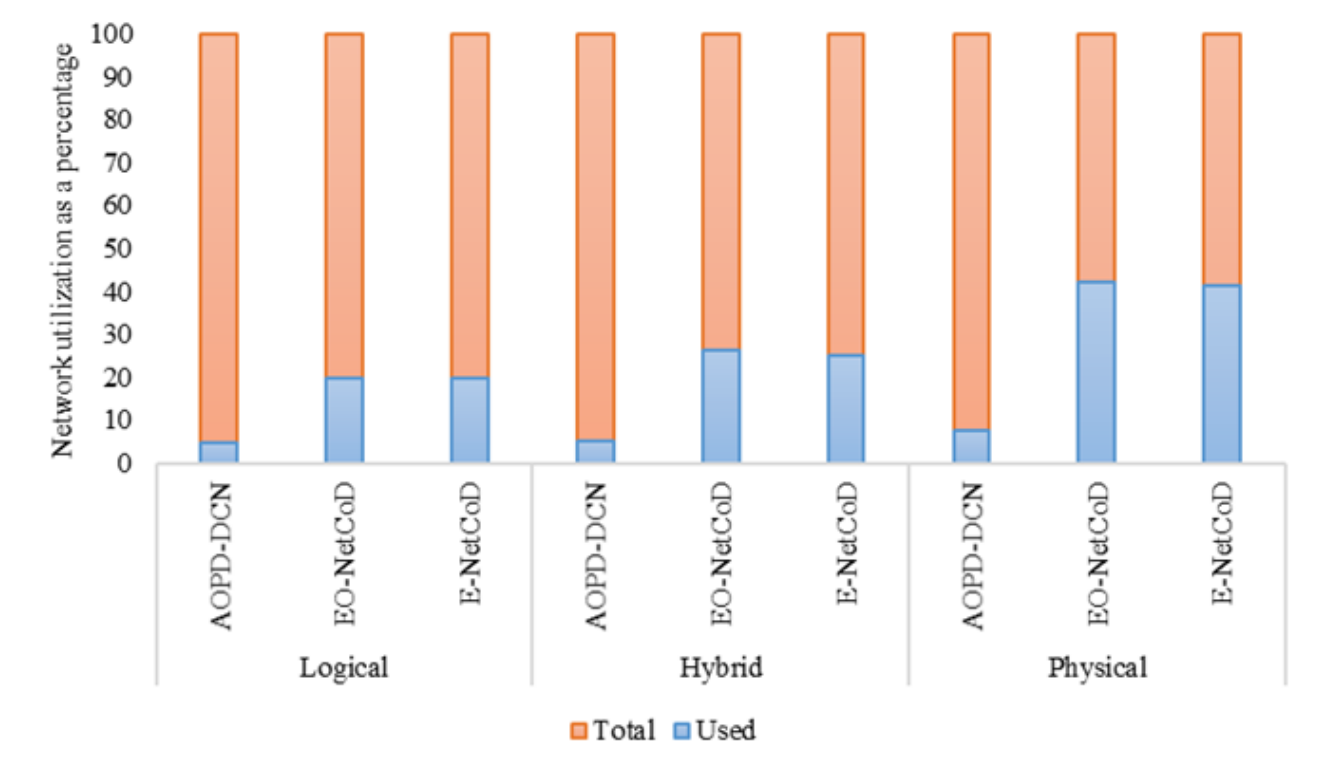

(b) Available network throughput utilization

Fig. 12. Available throughput of capacitated network in composable DC and corresponding throughput utilization.

Relative to AOPD-DCN, the maximum throughput achievable by both variants of NetCoD is significantly lower as shown in Fig. 12(a). However, it is important to remember that in AOPD-DCN, each compute node has a dedicated 400 Gbps interface to communicate directly with each co-rack compute node. In addition, each compute node in AOPD-DCN also has a total of 800 Gbps that is available to communicate via the inter-rack network. On the other hand, each compute node has a maximum of 800 Gbps to communicate with all nodes in the DC when both variants of NetCoD are considered. However, both variants of NetCoD achieve significantly higher utilization of the available network throughput as shown in Fig. 12(b). About 20% of E-NetCoD and EO-NetCoD available capacity is used when logical disaggregation is implemented in the composable rack as illustrated in Fig. 12(b). On the other hand, the higher throughput provided by AOPD-DCN significantly exceeds the practical need in the logically disaggregated DCs. Only about 5% of AOPD-DCN available throughput is used as illustrated in Fig. 12(b). Therefore, relative to the technically challenging generic design adopted for AOPD-DCN, the targeted design adopted for NetCoD achieves greater utilization (4 times greater) while delivering similar performance.

*2) Hybrid Disaggregation*

The results show that the general placement strategy observed under in the logically disaggregated rack-scale DC is implemented when hybrid disaggregation is employed in the rack-scale composable DC i.e., VM rejection remains discouraged as a balance between compute energy efficiency and network energy efficiency is found. However, relative to the zero-cost-and-un-capacitated network, total compute power consumption is marginally (below a percent) higher when AOPD-DCN, E-NetCoD or EO-NetCoD is employed. This is because maximum compute energy efficiency is not achieved when network cost and constraints are introduced. Given both compute and network constraints, marginal concessions in compute energy efficient are required to achieve an optimal result.

Relative to the zero-cost-and-un-capacitated network, the total compute power consumption obtained when AOPD-DCN is deployed in rack-scale composable DC that implements hybrid disaggregation is marginally (less that a percent) lower. A marginal increase in the TMPC is solely responsible for the marginal increase in total compute power consumption. Revisions in the placement of VMs' memory demands because of network constraints is responsible for the increase in TMPC. Although the same number of memory components are utilized, and the average active utilization of memory component increased as shown in Fig. 10(b), a less energy efficient memory component is highly utilized. Therefore, the TMPC increased accordingly as shown in Fig. 11(b). Relative to the zero-cost-and-un-capacitated network, the placement of VMs' CPU and storage demands is different when AOPD-DCN was deployed in the rack-scale DC. However, Fig. 11(a), Fig. 10(a), and Fig. 10(b) respectively show that the revised placement achieved the same TCPC and TSPC, equal number of active CPU and storage components and the same average active compute components utilization as zero-cost-and-un-capacitated network.

The placement of VMs obtained when E-NetCoD is deployed is a replica of the placement obtained when AOPD-DCN was deployed in the rack-scale composable DC. Hence, equal number of active compute components is obtained under both scenarios and the average utilization of active CPU, memory and storage components is equal under both scenarios.

Consequently, equal total compute power consumption (TCPC, TMPC and TSPC) is obtained when E-NetCoD is deployed to replace AODP-DCN in a rack-scale DC that implements hybrid disaggregation. Despite the equal TCPC, it is important to note that the resulting placement of CPU demands obtained under E-NetCoD is different from the placement of CPU demands obtained under the zero-cost-and-un-capacitated network.

The placement of VMs when EO-NetCoD implemented is comparable to the placement of VMs obtained when E-NetCoD is deployed in the composable DC. Equal TCPC and TSPC are obtained under both scenarios. This is because equal number of CPU and storage components are activated under both scenarios. The average utilization of active CPU and storage component obtained when EO-NetCoD is deployed is equal to the corresponding values obtained when E-NetCoD is implemented in the DC. Although, 8 RAM components are active under both scenarios, a slight variation in the placement of memory demands under EO-NetCoD is responsible for marginal increase in the TMPC relative to the corresponding value obtained under E-NetCoD. The marginally higher average utilization of active memory component under EO-NetCoD as shown in Fig. 10(b) is because a memory component with lower energy efficiency is highly utilized compared to when E-NetCoD is deployed in the composable DC.

The results obtained by solving the MILP model also show attempts to minimize TNPC when AOPD-DCN, E-NetCoD and EO-NetCoD topologies are employed in the DC. AOPD-DCN primarily employs zero-hop communication between intra-rack compute nodes to ensure that both TNFP and TNRP are minimized. However, in one instance where the capacity of the direct interface between nodes is limited, an additional path is established via the optical ToR switch to support intra-rack traffic. Note that it is more energy efficient to setup a light-path via the optical switch than it is to use another compute node as an intermediate node. This is because of the associated routing and forwarding power that will be consumed at the intermediate node. In the physical layer, direct light-paths are setup between each compute node and the gateway switch via the optical ToR switch to enable transmission of north-south traffic in both northbound and southbound directions.

As observed when AOPD-DCN is employed in the DC, direct light-paths are often established between communicating node pairs in the DC to carry traffic when EO-NetCOD is employed. However, multi-hop communication is also periodically used to transmit low data rate (mice) traffic to ensure optimal utilization of provisioned light-paths. Such low data rate (mice) traffic is piggybacked on other light-paths that are established to convey low-medium data rate traffic to an intermediate node. The intermediate node thereafter sets-up another lightpath to jointly forward the transiting mice traffic and its own traffic to a destination node. Multi-hop communication paths can be provisioned in both the logical and physical layers of the corresponding network topology. It is important to note that a large traffic demand between two nodes maybe divided into mice and elephant portions to ensure optimal utilization of the network. On the one hand, the mice portion maybe forwarded via multi-hop communication path to optimally utilize the network by maximizing the utilization of each active lightpath. On the other hand, the elephant portion of the divided traffic is forwarded via zero-hop communication path to reduce the SOAPC, TNRP and TNFP. This is because forwarding elephant traffic via multi-hop communication paths significantly increases TNPC.

E-NetCoD adopts a similar approach as EO-NetCoD to maximize the utilization of the active light-paths and to mitigate the impact of network constraints. When the E-NetCoD is employed in rack-scale composable DC, direct virtual links are setup between source and destination nodes of the traffic demand in the logical layer. However, in the physical layer, mice flows of the virtual layer are often forwarded via multi-hop communication path. Such flows are carried on a common lightpath with low-medium sized flows that are destined for the selected intermediate compute node. At the intermediate compute node, the transiting mice flow is piggybacked on a different lightpath that is setup to convey another low-medium sized flow originating at the intermediate compute node. This strategy is commonly observed for mice flows that originate from compute nodes that comprise of multiple CPU components. Such nodes are highly constrained when the CPU demand of multiple VMs are placed in them. Hence, multiple light-paths must be provisioned to convey inter-resource traffic to and from storage and memory components within the rack as well as north-south traffic to and from the inter-rack network in the DC. On the other hand, the elephant flows are usually forwarded via zero-hop communication path to reduce the SOAPC, TNRP and TNFP.

Generally, in the DC that implements hybrid disaggregation, AOPD-DCN has the lowest TNPC, as seen in Fig. 11(b), because multi-hop communication is reduced and AOPD-DCN does not require SOA switches. However, AOPD-DCN has inherent technical implementation challenges that must be addressed in a practical scenario. Even if such challenges are addressed, an implementation of AOPD-DCN will be grossly underutilized as shown in the Fig. 12(b). On the other hand, the utilization of the available throughput when both variants of NetCoD are deployed in a rack-scale DC that implements hybrid disaggregation is 5 times greater than the utilization obtain under AOPD-DCN under a similar scenario as shown in Fig. 12(b).

EO-NetCoD has lower TNPC compared to E-NetCoD, as seen in Fig. 11(b), because energy intensive electrical ToR switches (with relatively high load proportional PC and operational PC) are not used in EO-NetCoD. TOBP is the same under all topologies considered because a similar placement strategy ensures that maximum traffic is exchanged via the highly energy efficient onboard fabric under all topologies. This strategy helps to minimize TNPC. As expected, Fig. 11(b) shows that the TNFP of E-NetCoD is the highest because an electrical ToR switch is used. The TNFP of EO-NetCoD is higher than that of AOPD-DCN as shown in Fig. 11(b) because single-hop communication (via other compute nodes) is infrequently adopted to optimally utilize network capacity under EO-NetCoD. This leads to additional forwarding and routing power that are absent when AOPD-DCN is employed. Furthermore, both variants of NetCoD require SOA switches, which also introduce additional network power while AOPD-DCN does not.

For all network topologies considered, relative to when the DC was logically disaggregated, the TNPC is marginally higher in a DC that implements hybrid disaggregation as seen in Fig. 11(b). Relative to the logically disaggregated DC, the TNPC of a DC that implements hybrid disaggregation increased by 0.9%, 0.7% and 1.3% when AOPD-DCN, E-NetCoD and EO-NetCoD are employed, respectively. This is because the traffic in higher tiers of the network increases when hybrid disaggregation is employed. Consequently, both forwarding and routing power increase accordingly. Furthermore, more SOA switches are required in both variants of NetCoD when hybrid disaggregation is implemented in the composable DC. The relatively marginal increase in TNPC when hybrid disaggregation is implemented instead of logical disaggregation is because next generation energy efficiency values are adopted for different tiers of the network. Hence, the transmission of significantly higher volumes of traffic when hybrid disaggregation is implemented in the composable DC does not lead to drastic increase in TNPC. It is important to note that comparable compute power consumption is achieved when logical or hybrid disaggregation is adopted in a rack-scale composable DC as shown in Fig. 11(a).

*3) Physical Disaggregation*

Relative to the placement of VM resource demands under the zero-cost-and-un-capacitated network, the results show that introduction of network constraints and cost led to changes in the placement of VM demands under AOPD-DCN, E-NetCoD and EO-NetCoD topologies. The results obtained when AOPD-DCN is deployed show that a strategy, which gives higher priority to energy efficient utilization of active CPU components, is adopted to achieve optimal energy efficiency in the DC. This is because CPU components consume more power than other components in the DC. However, relative to the zero-cost-and-un-capacitated network, revisions in the placement of CPU resource demands when AOPD-DCN is deployed in a physically disaggregated DC achieved the same TCPC as shown in Fig. 11(a). Similarly, equal number of active CPU components and average active CPU component utilization are obtained under both scenarios as shown in Fig. 10.

The result also shows that the memory demand of VMs, which belong to a common IMCG, are placed in the same compute node when AOPD-DCN is implemented. Hence, physical disaggregation is leveraged to reduce and/or eliminate in-memory communication traffic in the network since such placement is more feasible under the physical disaggregation. Consequently, this minimizes forwarding power in the network and can also reduce multi-hop communication. In most situation, memory demands are placed in a manner that ensures that VMs in the same IMCG are placed in the same compute node. This minimizes in-memory communication in the composable DC. Relative to the zero-cost-and-un-capacitated network, the placement strategy adopted for memory resource demands is responsible for a marginal (less than 1%) rise in the TMPC under AOPD-DCN. Therefore, further highlighting the need for marginal concessions in compute energy efficiency to achieve optimal overall efficiency. A different storage demand placement under the AOPD-DCN achieved the same efficiency as the zero-cost-and-un-capacitated network because most storage demands are non-intensive as illustrated in Table IV. Hence, bin-packing storage demands for maximum energy efficiency is highly feasible while achieving optimal total efficiency. Consequently, only necessary, and minimal amount of storage components are activated when AOPD-DCN is employed in the physically disaggregated DC. The same number and type of storage components are activated under both AOPD-DCN and the zero-cost-and-un-capacitated network; and the average active storage component utilization is also equal under both scenarios as shown in Fig. 10. Consequently, the same TSPC is obtained under both scenarios as seen in Fig. 11(a).

When feasible, attempts are made to maximize network utilization under AOPD-DCN via coordinated placement of CPU, memory, and storage resource demand. Such coordination ensures that active light-paths are shared to improve their utilization. In the logical layer direct virtual links are setup between all communicating nodes pairs. Intra-rack traffic is often sent via direct point-to-point links between compute nodes in the rack. However, when the traffic demand between two nodes in the same rack exceeds the capacity of the direct link (i.e., 400 Gbps) that connects two compute nodes, additional light-paths are provisioned over the inter-rack network to supplement the point-to-point capacity in the intra-rack network. This strategy is common for compute nodes which hold multiple CPU components since the direct 400 Gbps link between compute nodes may be inadequate for very large or aggregated CPU-to-memory traffic in a composable DC. A direct virtual link (optical light path) is created between the gateway switch and each compute node that has north-south traffic in either northbound or southbound direction. The optical ToR switch serves as a transparent intermediate node between compute nodes and the gateway switch in the physical layer. It is also important to note that some provisioned direct lightpaths are poorly utilized when AOPD-DCN is employed. However, since, there is no penalty for poorly utilized lightpath under AOPD-DCN, this is an acceptable outcome.

The placement of VMs also shows that a VM is rejected when both variants of NetCoD are deployed in the physically disaggregated DC. Limited network capacity at compute nodes is responsible for such VM rejection. However, rejection is easily mitigated via the introduction of additional compute nodes in the rack. Moreover, compared to the small evaluation scenarios considered in this paper, typical DCs usually have over-provisioned hardware capacity to mitigate such rejection and to ensure that service level agreements at met. Hence, in practice such rejection is unlikely to occur. Although results under both variants of NetCoD also show that it is important to give high priority to energy efficiency of CPU component. However, given the limited number of CPU compute nodes considered, satisfying the CPU demand of all VMs under either variant of NetCoD while enforcing network constraints is infeasible. Hence, VM rejection is unavoidable. Furthermore, since a common cost of rejection is associate with all VMs considered, VM 6, which is known to have very high compute and network requirement as seen in Table IV, is rejected under E-NetCoD and EO-NetCoD.

Compared to the results obtained under E-NetCoD, the TMPC and TSPC obtained under EO-NetCoD are marginally higher while equal TCPC is obtained under both variants of NetCoD. Equal number of active CPU, memory and storage components were obtained under both scenarios and the average utilization of each resource type is comparable under both scenarios as illustrated in Fig. 10. Generally, the TCPC, TMPC and TSPC obtained under both variants of NetCoD are lower compared to similar values obtained under zero-cost-and-un-capacitated network and AOPD-DCN as shown in Fig. 11(a). VM rejection under both variants of NetCoD is responsible for this trend.

The results also demonstrate the importance of making strategic placement of compute demands to achieve optimal results while satisfying network constraints. A common strategy employed under both variants of NetCoD is to systematically place compute demands into the rack in a manner that ensures the satisfaction of network constraints. Compute demand placement also attempts to minimize the number of communicating node pairs in the rack. However, this can be difficult, especially for compute nodes with multiple CPU resource in a physically disaggregated DC. Such compute nodes can host CPU demands of multiple VMs and must support the aggregated CPU-to-memory, CPU-to-storage, and CPU-to-gateway traffic in both directions for all VMs placed in that compute node. Multi-hop communication is employed to mitigate stringent network constraints by ensuring that data traffic is optimally aggregated on provisioned light-paths. Multi-hop communication is adopted in two instances as observed when hybrid disaggregation was implemented in the composable DC.

In the first instance, mice traffic such as CPU-to-storage and CPU-to-gateway traffic in the rack originating from compute nodes are often forwarded via multi-hop communication path. Such mice flows are aggregated and sent over a lightpath established to convey low-medium data rate traffic to intermediate compute nodes. The intermediate compute node receives and processes the traffic that is destined for it and forwards the transiting traffic to the next hop on the multi-hop communication path. Multi-hop communication can be set up in both logical and physical layers of the network.

In another instance, large-sized and/or aggregated CPU-to-memory traffic from a compute node can be divided into multiple streams to be forwarded on optical light-paths using single wavelength data rate (100 Gbps) as a divisor. On the one hand, larger (elephant) portions of such traffic are thereafter transmitted via zero-hop paths to minimize network power consumption by maximizing active light path utilization. On the other hand, small (mice) portion of such traffic are forwarded over multi-hop paths to maximize network utilization and to mitigate the impact of stringent network constraints. However, the adoption of multi-hop communication in this manner can lead to performance degradation since CPU-to-memory traffic is known to be latency sensitive.

The results obtained when E-NetCoD is employed in the physically disaggregated DC show that direct virtual links are created between all communicating node pairs; hence, traffic is not relayed in the virtual layer of the network. In the physical layer, high data rate (elephant) flows of each virtual link are sent via direct physical links while smaller (mice) flows of each virtual link are piggybacked on established light-paths between compute nodes. This helps to promote greater lightpath utilization. However, a practical implementation must ensure that latency sensitive traffic, such as CPU-memory communication, are sent over minimal number of hops. Generally, southbound traffic from the gateway switch to compute nodes is transmitted over direct light-paths setup up from the gateway switch via the optical ToR to each compute node that will receive such traffic. The results show that the same routing and forwarding strategies implemented when E-NetCoD is deployed in physically disaggregated DC are also employed when EO-NetCoD is deployed in the DC.

Relative to results obtain when logical or hybrid disaggregation is adopted in the single rack, result obtain when physical disaggregation is implemented show that TNPC increases marginally as seen Fig. 11(b). Compared to the power consumption of a logically disaggregated DC that employed AOPD-DCN, E-NetCoD and EO-NetCoD, the TNPC increased by 3.7%, 2.6% and 4.4% respectively when the same DC is physically disaggregated. Similar comparison between a DC that implements hybrid disaggregation and physical disaggregation shows that the TNPC of AOPD-DCN, E-NetCoD and EO-NetCoD increased by 2.8%, 1.8% and 3.1% respectively when a DC is physically disaggregated. Increase in TNRP and TNFP due to increase in network traffic traversing the intra-rack network is primarily responsible for the observed results under all network topologies considered. Increase in the SOAPC when physical disaggregation is deployed in the rack is also contributed to the increase in TNPC when both variants of NetCoD are employed.

Compared to both variants of NetCoD, the TNFP and TNRP obtained when AOPD-DCN is deployed in a rack that employs physical disaggregation is higher. This is because all VMs are provisioned when AOPD-DCN was deployed while a VM is rejected when both variants of NetCoD were deploy in the DC. A similar reason justifies the lower TOBP under both variants of NetCoD relative to AOPD-DCN. Fig. 12(b) shows the utilization of the available network throughput of both variants of NetCoD is over 5 times greater than that of AOPD-DCN when a physically disaggregated rack-scale DC is considered. Hence, it is expected that the cost of implementing AOPD-DCN will outweigh the practical benefits that are derived. On the other hand, both variants of NetCoD can provide the required network capacity to support physical disaggregation using a more specific and practical design relative to the general-purpose design adopted for AOPD-DCN. However, great intelligence is required to achieve optimal efficiency.

## VII. Conclusions

In this paper, we described two variants of the network for composable DCs (NetCoD). In contrast to the general-purpose design employed when mesh physical topology is employed in the intra-rack networks of composable DCs, a more targeted design is adopted in NetCoD. The targeted design leveraged optical technologies and techniques to reduce complexity and cost. Using a MILP model for capacitated networks, we compare the performance of both variants of NetCoD to the performance of a reference network topology in a DC with multiple racks. The electrical-optical variant of NetCoD achieved comparable network energy efficiency as the reference topology. But the energy efficiency of the all-electrical variant of NetCoD is relatively lower. Subsequently, we extended the MILP model formulated for the capacitated networks, to consider energy efficient placement of VMs in a rack-scale composable DC that implemented logical, hybrid and physical disaggregation. Relative to the reference topology, the results showed that both variants of NetCoD achieved similar compute energy efficiency under all forms of disaggregation. The range of scenarios considered highlighted various strategies that can be deployed in practical implementations of either variant of NetCoD to improve overall energy efficiency in composable DCs while satisfying both compute and network constraints. Under all network topologies considered, a logically disaggregated DC achieved the best results. Across all network topologies evaluated in this paper, the average increase in TNPC is 1% and 3.6% when hybrid and physical disaggregation are respectively implemented instead of logical disaggregation in the small evaluation scenario considered. Additionally, the utilization of available network throughput by both variants of NetCoD exceeds that of reference topology by 4 – 5 times under the different forms of disaggregation in the rack-scale composable DC. A limitation of the MILP optimization methodology adopted to conduct this study is the need for high computing power for a long duration to obtain an optimal solution. This is because a detailed MILP model is needed to effectively represent the problems considered. Hence, a high number of binary variables are required to represent the placement of VM compute demands and network wavelength assignment. This prevented the consideration of multi-rack and multi-cluster composable DCs. However, the single rack scenario considered in Section VI effectively validated and demonstrated the efficacy of the targeted design proposed for composable DCNs in this paper.

**Opeyemi O. Ajibola** received the B.Sc. degree (High Hons.) in Electrical and Electronic Engineering from Eastern Mediterranean University, Famagusta, North – Cyprus, in 2011 and the M.Sc. degree (with distinction) in Digital Communications Networks from University of Leeds, Leeds, UK in 2015. He is currently working towards the PhD degree in the School of Electronic and Electrical Engineering, University of Leeds, Leeds, UK. From 2012 to 2013, he was a Wireless Solution Sales Engineer with Huawei Technologies, Abuja, Nigeria. In 2014, he joined Federal University Oye-Ekiti, Ekiti State, Nigeria as a graduate assistant. His research interests include composable data centers, energy efficient data centers and communication networks, energy efficient cloud and fog/edge computing and the Internet of Things.

**Taisir E. H. El-Gorashi** received the B.S. degree (first-class Hons.) in Electrical and Electronic Engineering from the University of Khartoum, Khartoum, Sudan, in 2004, the M.Sc. degree (with distinction) in Photonic and Communication Systems from the University of Wales, Swansea, UK, in 2005, and the PhD degree in Optical Networking from the University of Leeds, Leeds, UK, in 2010. She is currently a Lecturer in optical networks in the School of Electronic and Electrical Engineering, University of Leeds. Previously, she held a Postdoctoral Research post at the University of Leeds (2010– 2014), where she focused on the energy efficiency of optical networks investigating the use of renewable energy in core networks, green IP over WDM networks with datacenters, energy efficient physical topology design, energy efficiency of content distribution networks, distributed cloud computing, network virtualization and big data. In 2012, she was a BT Research Fellow, where she developed energy efficient hybrid wireless-optical broadband access networks and explored the dynamics of TV viewing behavior and program popularity. The energy efficiency techniques developed during her postdoctoral research contributed 3 out of the 8 carefully chosen core network energy efficiency improvement measures recommended by the GreenTouch consortium for every operator network worldwide. Her work led to several invited talks at GreenTouch, Bell Labs, Optical Network Design and Modelling conference, Optical Fiber Communications conference, International Conference on Computer Communications, EU Future Internet Assembly, IEEE Sustainable ICT Summit and IEEE 5G World Forum and collaboration with Nokia and Huawei.

**Professor Jaafar Elmirghani** is Fellow of IEEE, Fellow of the IET, Fellow of the Institute of Physics and is the Director of the Institute of Communication and Power Networks and Professor of Communication Networks and Systems within the School of Electronic and Electrical Engineering, University of Leeds, UK. He joined Leeds in 2007 having been full professor and chair in Optical Communications at the University of Wales Swansea 2000-2007.

He received the BSc in Electrical Engineering, First Class Honours from the University of Khartoum in 1989 and was awarded all 4 prizes in the department for academic distinction. He received the PhD in the synchronization of optical systems and optical receiver design from the University of Huddersfield UK in 1994 and the DSc in Communication Systems and Networks from University of Leeds, UK, in 2012. He co-authored Photonic Switching Technology: Systems and Networks, (Wiley) and has published over 550 papers.

He was Chairman of the IEEE UK and RI Communications Chapter and was Chairman of IEEE Comsoc Transmission Access and Optical Systems Committee and Chairman of IEEE Comsoc Signal Processing and Communication Electronics (SPCE) Committee. He was a member of IEEE ComSoc Technical Activities Council' (TAC), is an editor of IEEE Communications Magazine and is and has been on the technical program committee of 41 IEEE ICC/GLOBECOM conferences between 1995 and 2020 including 19 times as Symposium Chair. He was founding Chair of the Advanced Signal Processing for Communication Symposium which started at IEEE GLOBECOM'99 and has continued since at every ICC and GLOBECOM. Prof. Elmirghani was also founding Chair of the first IEEE ICC/GLOBECOM optical symposium at GLOBECOM'00, the Future Photonic Network Technologies, Architectures and Protocols Symposium. He chaired this Symposium, which continues to date. He was the founding chair of the first Green Track at ICC/GLOBECOM at GLOBECOM 2011, and is Chair of the IEEE Sustainable ICT Initiative, a pan IEEE Societies Initiative responsible for Green ICT activities across IEEE, 2012-present. He has given over 90 invited and keynote talks over the past 15 years.

He received the IEEE Communications Society 2005 Hal Sobol award for exemplary service to meetings and conferences, the IEEE Communications Society 2005 Chapter Achievement award, the University of Wales Swansea inaugural 'Outstanding Research Achievement Award', 2006, the IEEE Communications Society Signal Processing and Communication Electronics outstanding service award, 2009, a best paper award at IEEE ICC'2013, the IEEE Comsoc Transmission Access and Optical Systems outstanding Service award 2015 in recognition of "Leadership and Contributions to the Area of Green Communications", the GreenTouch 1000x award in 2015 for "pioneering research contributions to the field of energy efficiency in telecommunications", the IET 2016 Premium Award for best paper in IET Optoelectronics, shared the 2016 Edison Award in the collective disruption category with a team of 6 from GreenTouch for their joint work on the GreenMeter, the IEEE Communications Society Transmission, Access and Optical Systems technical committee 2020 Outstanding Technical Achievement Award for outstanding contributions to the "energy efficiency of optical communication systems and networks". He was named among the top 2% of scientists in the world by citations in 2019 in Elsevier Scopus, Stanford University database which includes the top 2% of scientists in 22 scientific disciplines and 176 sub-domains. He was elected Fellow of IEEE for "Contributions to Energy-Efficient Communications," 2021.

He is currently an Area Editor of IEEE Journal on Selected Areas in Communications series on Machine Learning for Communications, an editor of IEEE Journal of Lightwave Technology, IET Optoelectronics and Journal of Optical Communications, and was editor of IEEE Communications Surveys and Tutorials and IEEE Journal on Selected Areas in Communications series on Green Communications and Networking. He was Co-Chair of the GreenTouch Wired, Core and Access Networks Working Group, an adviser to the Commonwealth Scholarship Commission, member of the Royal Society International Joint Projects Panel and member of the Engineering and Physical Sciences Research Council (EPSRC) College.

He has been awarded in excess of £30 million in grants to date from EPSRC, the EU and industry and has held prestigious fellowships funded by the Royal Society and by BT. He was an IEEE Comsoc Distinguished Lecturer 2013-2016. He was PI of the £6m EPSRC Intelligent Energy Aware Networks (INTERNET) Programme Grant, 2010-2016 and is currently PI of the EPSRC £6.6m Terabit Bidirectional Multi-user Optical Wireless System (TOWS) for 6G LiFi, 2019-2024. He leads a number of research projects and has research interests in communication networks, wireless and optical communication systems.